\documentclass[11pt]{article}

\title{First document}

\usepackage[margin=1.0in]{geometry}

\usepackage{graphics}
\usepackage{url}
\usepackage{caption}
\usepackage{amsmath}
\usepackage{hyperref}
\usepackage{cite}
\usepackage{slashed}
\usepackage[colorinlistoftodos, shadow]{todonotes}
\usepackage{titlesec}
\usepackage{feynarts}
\usepackage{gensymb}
\usepackage{ulem}
\usepackage{amssymb}

\def\refeq#1{\mbox{(\ref{#1})}}
\def\reffi#1{\mbox{Fig.~\ref{#1}}}
\def\reffis#1{\mbox{Figs.~\ref{#1}}}
\def\refta#1{\mbox{Table~\ref{#1}}}

\def\refse#1{\mbox{Section~\ref{#1}}}

\def\citere#1{\mbox{Ref.~\cite{#1}}}
\def\citeres#1{\mbox{Refs.~\cite{#1}}}

\def\al{\alpha}

\def\ga{\gamma}
\def\de{\delta}

\def\si{\sigma}

\def\La{\Lambda}

\newcommand{\GeV}{\unskip\,\mathrm{GeV}}
\newcommand{\MeV}{\unskip\,\mathrm{MeV}}
\newcommand{\TeV}{\unskip\,\mathrm{TeV}}
\newcommand{\fba}{\unskip\,\mathrm{fb}}

\def\mathswitch#1{\relax\ifmmode#1\else$#1$\fi}
\def\mathswitchr#1{\relax\ifmmode{\mathrm{#1}}\else$\mathrm{#1}$\fi}
\def\mathswitchit#1{\relax\ifmmode{#1}\else$#1$\fi}

\newcommand{\Pu}{\mathswitchr u}
\newcommand{\Pd}{\mathswitchr d}
\newcommand{\Ps}{\mathswitchr s}
\newcommand{\Pc}{\mathswitchr c}
\newcommand{\Pb}{\mathswitchr b}
\newcommand{\Pt}{\mathswitchr t}
\newcommand{\Pp}{\mathswitchr p}

\newcommand{\Pq}{\mathswitchit q}
\newcommand{\Pl}{\mathswitch l}
\newcommand{\Plp}{\mathswitch l^+}
\newcommand{\Plm}{\mathswitch l^-}
\newcommand{\Pe}{\mathswitchr e}

\newcommand{\PW}{\mathswitchr W}
\newcommand{\PZ}{\mathswitchr Z}

\newcommand{\Pg}{g}
\newcommand{\PH}{\mathswitchr H}

\newcommand{\Pnl}{\mathswitch {\nu_\Pl}}
\newcommand{\Pne}{\mathswitch \nu_{\mathrm{e}}}
\newcommand{\Panl}{\mathswitch {\bar\nu_\Pl}}

\newcommand{\Pnmu}{\mathswitch \nu_\mu}
\newcommand{\Pnta}{\mathswitch \nu_\tau}

\newcommand{\MW}{\mathswitch {M_\PW}}

\newcommand{\MZ}{\mathswitch {M_\PZ}}

\newcommand{\GF}{\mathswitch {G_\mu}}

\newcommand{\ri}{{\mathrm{i}}}

\newcommand{\EW}{{\mathrm{EW}}}
\newcommand{\QCD}{{\mathrm{QCD}}}

\newcommand{\phot}{{\mathrm{phot}}}

\newcommand{\OS}{{\mathrm{OS}}}

\newcommand{\veto}{{\mathrm{veto}}}

\newcommand{\NLO}{{\mathrm{NLO}}}

\newcommand{\abs}[1]{\left|#1\right|}

\newcommand{\klam}[1]{\left(#1\right)}

\newcommand{\mr}[1]{\mathrm{#1}}

\newcommand{\Pbbar}{{\bar{\Pb}}}

\newcommand{\Pqbar}{{\bar{\Pq}}}

\newcommand{\phz}{\phantom{(0)}}

\allowdisplaybreaks[1]

\newcommand{\change}[1]{{#1}}

\def\draftdate{\relax}
\def\mda{\relax}
\def\mua{\relax}
\def\mla{\relax}
\def\draft{
\def\thtystars{******************************}
\def\sixtystars{\thtystars\thtystars}
\typeout{}
\typeout{\sixtystars**}
\typeout{* Draft mode!
         For final version remove \protect\draft\space in source file *}
\typeout{\sixtystars**}
\typeout{}
\def\draftdate{\today}
\def\mua{\marginpar[\boldmath\hfil$\uparrow$]%
                   {\boldmath$\uparrow$\hfil}%
                    \typeout{marginpar: $\uparrow$}\ignorespaces}
\def\mda{\marginpar[\boldmath\hfil$\downarrow$]%
                   {\boldmath$\downarrow$\hfil}%
                    \typeout{marginpar: $\downarrow$}\ignorespaces}
\def\mla{\marginpar[\boldmath\hfil$\rightarrow$]%
                   {\boldmath$\leftarrow $\hfil}%
                    \typeout{marginpar: $\leftrightarrow$}\ignorespaces}
\def\Mua{\marginpar[\boldmath\hfil$\Uparrow$]%
                   {\boldmath$\Uparrow$\hfil}%
                    \typeout{marginpar: $\uparrow$}\ignorespaces}
\def\Mda{\marginpar[\boldmath\hfil$\Downarrow$]%
                   {\boldmath$\Downarrow$\hfil}%
                    \typeout{marginpar: $\downarrow$}\ignorespaces}
\def\Mla{\marginpar[\boldmath\hfil$\Rightarrow$]%
                   {\boldmath$\Leftarrow $\hfil}%
                    \typeout{marginpar: $\leftrightarrow$}\ignorespaces}
\overfullrule 5pt
\oddsidemargin -15mm
\marginparwidth 29mm
}

\titleformat*{\subsubsection}{\large\it}

\numberwithin{equation}{section}

\begin{document}

\thispagestyle{empty}
\def\thefootnote{\fnsymbol{footnote}}
\setcounter{footnote}{1}
\null
\draftdate\hfill FR-PHENO-2015-012

\vfill
\begin{center}
  {\Large \boldmath{\bf NLO QCD and electroweak corrections to $\PZ+\gamma$
      production \\[.5em] with leptonic Z-boson decays}
\par} \vskip 2.5em
{\large
{\sc Ansgar Denner$^{1}$, Stefan Dittmaier$^{2}$, 
     Markus Hecht$^{2}$, Christian Pasold$^{1}$
}\\[2ex]
{\normalsize \it 
$^1$Julius-Maximilians-Universit\"at W\"urzburg, 
Institut f\"ur Theoretische Physik und Astrophysik, \\
D-97074 W\"urzburg, Germany
}\\[2ex]
{\normalsize \it 
$^2$Albert-Ludwigs-Universit\"at Freiburg, Physikalisches Institut, \\
D-79104 Freiburg, Germany
}\\[2ex]
}
\par \vskip 1em
\end{center}\par
\vskip .0cm \vfill {\bf Abstract:} 
\par 
The next-to-leading-order electroweak corrections to $\Pp\Pp\to
l^+l^-/\bar\nu\nu+\gamma+X$ production, including all off-shell
effects of intermediate Z~bosons in the complex-mass scheme, are
calculated for LHC energies, revealing the typically expected large
corrections of tens of percent in the TeV range.  Contributions from
quark--photon and photon--photon initial states are taken into account
as well, but their impact is found to be moderate or small.  Moreover,
the known next-to-leading-order QCD corrections are reproduced.  In
order to separate hard photons from jets, both a quark-to-photon
fragmentation function \'a la Glover/Morgan and Frixione's cone
isolation are employed.  The calculation is available in the form of
Monte Carlo programs allowing for the evaluation of arbitrary
differential cross sections.  Predictions for integrated cross
sections are presented for the LHC at $7\TeV$, $8\TeV$, and $14\TeV$,
and differential distributions are discussed at $14\TeV$ for bare
muons and dressed leptons.  Finally, we consider the impact of
anomalous $\PZ\PZ\gamma$ and $\PZ\gamma\gamma$ couplings.

\par
\vskip 1cm
\noindent
October 2015
\par
\null
\setcounter{page}{0}
\clearpage
\def\thefootnote{\arabic{footnote}}
\setcounter{footnote}{0}

\section{Introduction}
\label{se:intro}

The production of a photon with a leptonically decaying Z~boson
represents an important process class at hadron colliders such as the
Tevatron and the LHC, both as precision test ground of the Standard
Model (SM) and as probe for new-physics effects.  The investigation of
charged lepton pairs at intermediate energy scales with an additional
photon is part of the high-precision analysis of inclusive Z-boson
production.  Moreover, the production of a photon and a charged lepton
pair is the main background to the search for the Higgs-boson decay
into a photon and a Z~boson, which can only be measured if the
theoretical prediction for the background is well under control
\cite{Djouadi:1991tka,Dawson:1990zj,Spira:1995rr,Djouadi:1996yq}.  At
high energies $\PZ+\gamma$ production develops a strong sensitivity to
potentially existing photon--Z-boson couplings ($\PZ\PZ\gamma$,
$\PZ\gamma\gamma$) which are absent in the SM as elementary
interactions, so that non-standard $\PZ\PZ\gamma$ and
$\PZ\gamma\gamma$ couplings can be constrained by investigating
$\PZ+\gamma$ final states.  Such constraints were already reported by
the Tevatron experiments~\cite{Aaltonen:2011zc,Abazov:2011qp} and
further tightened by the LHC experiments
ATLAS~\cite{Aad:2013izg,Aad:2014fha} and
CMS~\cite{Chatrchyan:2013fya,Khachatryan:2015kea}.  If the Z~boson
decays invisibly into a neutrino pair, the experimental signature is
mono-photon production with missing transverse energy, a signal that
is particularly interesting in many exotic new-physics models (see,
e.g.,
\citeres{Fox:2011pm,Belanger:2012mk,Gabrielli:2014oya,Maltoni:2015twa}).
Searches for such signals were both carried out at the
Tevatron~\cite{Aaltonen:2008hh,Abazov:2008kp} and the
LHC~\cite{Aad:2014tda,Khachatryan:2014rwa}.  In none of the
experimental analyses of $\PZ+\gamma$ production any signs of new
physics have been seen so far.  In order to carry on those analyses at
run~2 of the LHC with higher energy and luminosity, theoretical
predictions have to be pushed to a high level of precision, aiming at
uncertainties at the level of few percent.

The first calculations for $\PZ+\gamma$ production were performed at
leading order (LO) in \citere{Renard:1981es} in 1981.  Subsequently
next-to-leading-order (NLO) QCD corrections were calculated for
on-shell (stable) $\PZ$~bosons in \citere{Ohnemus:1992jn} and extended
to include leptonic decays in the narrow-width approximation and
anomalous couplings in \citeres{Ohnemus:1994qp,Baur:1997kz}.  A Monte
Carlo program for $\PZ+\gamma$ (and $\PW+\gamma$) production at NLO
QCD was presented in \citere{DeFlorian:2000sg} using amplitudes from
\citere{Dixon:1998py}, where the leptonic decays of the W/Z~bosons are
treated in the narrow-width approximation, while the spin information
is retained via decay-angle correlations. In the same approximation
the NLO QCD corrections to $\PZ+\gamma$ production are also included
in the publically available program MCFM \cite{Campbell:2011bn}.
Since the NLO QCD corrections are of the order of $50\%$, the NNLO QCD
corrections were expected to be sizeable.  Based on a scale-variation
analysis in \citere{Campbell:2011bn} they were estimated to be of the
order of $5\%$.  Since 2013 the NNLO QCD corrections are available,
and predictions for the LHC were published in
\change{\citere{Grazzini:2013bna,Grazzini:2015nwa}}, revealing a
residual scale dependence of only $\sim2\%$ for the integrated cross
section.
\change{The NNLO QCD predictions, in particular, include contributions
from the loop-induced gluon-fusion process $\Pg\Pg\to\PZ+\gamma$,
which was calculated in the approximation of stable Z~bosons
already a long time ago~\cite{vanderBij:1988fb}.}

It is well known that EW corrections can cause sizeable effects at high 
energies above the EW scale due to the presence of logarithmically 
enhanced contributions, so-called Sudakov (and subleading) logarithms 
\cite{Beenakker:1993tt,Beccaria:1998qe,Ciafaloni:1998xg,Kuhn:1999de,Fadin:1999bq,Denner:2000jv}.
EW corrections to $\PZ+\gamma$ production at hadron colliders
have been presented for on-shell Z~bosons in \citere{Hollik:2004tm}.
Shortly after, in \citere{Accomando:2005ra} the EW corrections to $\PZ+\gamma$ (and $\PW+\gamma$)
production have been calculated, including the decay of the massive 
vector bosons in pole approximation.

In this paper we push the existing calculations of EW corrections for
$\PZ+\gamma$ to the level of complete NLO EW calculations for the full
off-shell processes $\Pp\Pp\to\Plp\Plm/\bar\nu\nu\ga+X$, including all
partonic channels ($\Pq\gamma$ and $\gamma\gamma$) with initial-state
(IS) photons.  The NLO QCD corrections are rederived as well.  In
order to attribute collinear photon--jet configurations either to
$\PZ+\gamma$ or $\PZ+\mathrm{jet}$ production, we alternatively employ
a quark-to-photon fragmentation function \'a la Glover and
Morgan~\cite{Glover:1993qtp,Glover:1994th} or Frixione's cone
isolation~\cite{Frixione:1998jh}.

The paper is organized as follows: Section~\ref{se:details} 
briefly describes the setup and techniques of our calculation, referring to
the more detailed discussion~\cite{Denner:2014bna} 
of $\PW+\gamma$ production as much as possible, and
contains a survey of the calculated corrections.
In \refse{se:numres} we discuss our numerical results on total and
differential cross sections, both in the SM and including effects of anomalous
$\PZ\PZ\gamma$ and $\PZ\gamma\gamma$ couplings.
Finally, our conclusions are given in \refse{se:concl}.

\section{Details of the calculation}
\label{se:details} 

The calculation of NLO corrections to $\PZ+\gamma$ production follows
the methods described in Section~2 of \citere{Denner:2014bna} for
$\PW+\gamma$ production. In this section we focus on the differences
compared to that paper.

\subsection{General setup}
\label{se:setup}

The production of a leptonically decaying $\PZ$ boson in association with a hard photon includes two 
different final states. If the $\PZ$ boson decays into two charged leptons the LO partonic process 
reads
\begin{equation}
 \Pq_{i} \Pqbar_{i} \to \Plp \Plm \, \gamma \, ,
 \label{eq:LO_zgamma_ll}
\end{equation}
and if the $\PZ$ boson decays into two neutrinos it is
\begin{equation}
 \Pq_{i} \Pqbar_{i} \to \Panl \Pnl \, \gamma \, ,
 \label{eq:LO_zgamma_nn}
\end{equation}
where $\Pq_{i} = \Pu, \Pd, \Ps, \Pc, \Pb$ denotes any light quark. The
corresponding LO Feynman diagrams are shown in
\reffis{fi:born_zgamma_ll} and \ref{fi:born_zgamma_nn}. While for the
process defined in Eq.~\refeq{eq:LO_zgamma_ll} we assume $\Pl = \Pe$
or $\mu$, the neutrino process includes three families of neutrinos
$\Pnl = \Pne, \Pnmu, \Pnta$. For the process where the $\PZ$ boson
decays into two charged leptons we present results for one single
family of final-state (FS) leptons, for the process with neutrinos in
the final state we sum the cross sections over all three flavours.

\begin{figure}
  \begin{center}
  \tabcolsep 5pt
    \begin{tabular}{cccc}
      \input{plots/feynman_diagrams/born_zgamma_ll/born03.tex} &
      \input{plots/feynman_diagrams/born_zgamma_ll/born04.tex} &
      \input{plots/feynman_diagrams/born_zgamma_ll/born01.tex} &
      \input{plots/feynman_diagrams/born_zgamma_ll/born02.tex} 
    \end{tabular}
  \end{center}
  \vspace*{-2em}
  \caption{LO Feynman diagrams for the partonic process $\Pq_{i} \Pqbar_{i} \to \Plp \Plm 
           \, \gamma$.}
  \label{fi:born_zgamma_ll}
\end{figure}

\begin{figure}
  \begin{center}
  \tabcolsep 5pt
    \begin{tabular}{cc}
      \input{plots/feynman_diagrams/born_zgamma_nn/born01.tex} &
      \input{plots/feynman_diagrams/born_zgamma_nn/born02.tex}
    \end{tabular}
  \end{center}
  \vspace*{-2em}
  \caption{LO Feynman diagrams for the partonic process $\Pq_{i} \Pqbar_{i} \to \Panl \Pnl 
           \, \gamma$.}
  \label{fi:born_zgamma_nn}
\end{figure}

\begin{figure}
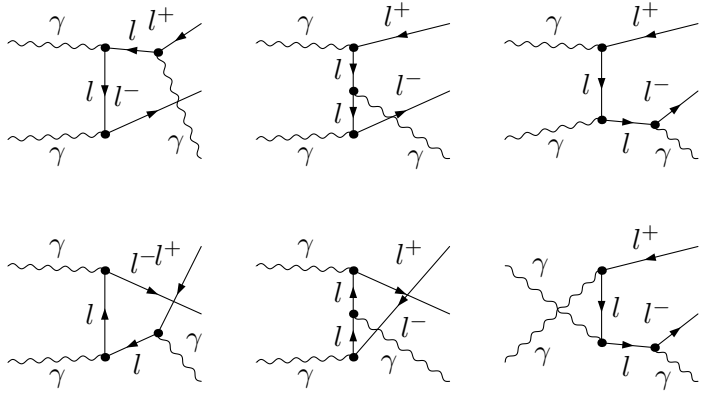

  \begin{center}
  \tabcolsep 5pt
    \begin{tabular}{ccc}
      \input{plots/feynman_diagrams/phph/phph06.tex} &
      \input{plots/feynman_diagrams/phph/phph01.tex} &
      \input{plots/feynman_diagrams/phph/phph02.tex} \\
      \input{plots/feynman_diagrams/phph/phph05.tex} &
      \input{plots/feynman_diagrams/phph/phph03.tex} &
      \input{plots/feynman_diagrams/phph/phph04.tex} 
    \end{tabular}
  \end{center}
  \vspace*{-2em}
  \caption{LO Feynman diagrams for the partonic process $\gamma \gamma \to \Plp \Plm  \gamma$.}
  \label{fi:phph}
\end{figure}

At LO the final state in Eq.~\refeq{eq:LO_zgamma_ll} can also be
produced via
\begin{equation}
 \gamma \, \gamma \to \Plp \Plm \, \gamma \, ,
 \label{eq:phph_zgamma_ll}
\end{equation}
which is a pure QED process and does not include any intermediate
vector boson. The corresponding LO Feynman diagrams 
are shown in \reffi{fi:phph}. 
Owing to the two photons in the initial state the 
partonic cross section is convoluted 
two times with the very small 
photon PDFs, so that the contribution to the pp 
cross section is expected to be small. For this reason 
we give results for its contribution separately and do not consider NLO EW 
corrections to this LO process. Since this process only contains charged 
leptons as intermediate particles there are no QCD corrections at NLO.

We choose to combine QCD and EW corrections to the quark--antiquark-induced channels using the naive 
product of the relative correction factors, 
while the quark--photon and the photon--photon contributions are added to the corrected 
$\Pq\overline{\Pq}$-induced cross section,
\begin{align}\label{eq:naive-product}
  \sigma^{\mr{NLO}}&= \sigma^{\mr{LO}}\left[\klam{1+\delta_{\QCD}}
                                  \klam{1+\delta_{\EW,\Pq\overline{\Pq}}}+\delta_{\EW, \Pq\gamma}+\left(\de_{\gamma\gamma}\right)\right]
\nonumber
\\
  &= \sigma^{\mr{NLO\,QCD}} \klam{1+\delta_{\EW,\Pq\overline{\Pq}}}+\Delta\sigma^{\NLO\,\EW}_{\Pq\gamma}+\left(\Delta\sigma_{\gamma\gamma}\right),
\end{align}
where the relative QCD, EW, and photon-induced corrections are defined by
\begin{align}
 \de_{\QCD} & = \frac{\si^{\NLO\,\QCD}-\si^{\mr{LO}}}{\si^{\mr{LO}}} , &
 \de_{\EW, \Pq\overline{\Pq}} & = \frac{\Delta\sigma^{\NLO\,\EW}_{\Pq\overline\Pq}}{\si^{0}} , \nonumber\\
 \de_{\EW, \Pq\gamma} & = \frac{\Delta\sigma^{\NLO\,\EW}_{\Pq\gamma}}{\si^{\mr{LO}}} , &
 \de_{\gamma\gamma} & = \frac{\Delta\sigma_{\gamma\gamma}}{\si^{\mr{LO}}} ,
\label{eq:relcor}
\end{align}
respectively. We have put the photon--photon channel in parentheses to
indicate that this channel does not contribute to neutrino production.
While the relative QCD and photon-induced corrections are normalized
to the LO cross section $\si^{\mr{LO}}$, calculated with LO PDFs, the
quark--antiquark-induced EW corrections are normalized to the LO cross
section $\si^{0}$, calculated with NLO PDFs. By this definition,
$K_{\QCD}=1+\delta_{\QCD}$ is the standard QCD $K$ factor, and the
relative quark--antiquark-induced EW corrections $\de_{\EW,
  \Pq\overline{\Pq}}$ are practically independent of the PDF set. In
case of $\PZ+\gamma$ production the purely weak $\de_{\mr{weak},
  \Pq\overline{\Pq}}$ and the photonic corrections $\de_{\mr{phot},
  \Pq\overline{\Pq}}$ can be separated in a gauge-independent way
\begin{align}
\label{eq:delta_qed_weak}
  \de_{\EW, \Pq\overline{\Pq}}= \de_{\mr{weak}, \Pq\overline{\Pq}}+\de_{\mr{phot}, \Pq\overline{\Pq}}. 
\end{align}
By definition, the photonic corrections comprise all diagrams with
photon exchange between fermi\-ons in a loop, the corresponding
counterterm contributions, and all photon emission effects.  All
remaining EW corrections to the $\Pq\bar\Pq$~channels furnish the weak
corrections.  Where appropriate we show the weak corrections and the
photonic corrections separately or we show the weak corrections
additionally to the EW corrections.

Note that the combination \refeq{eq:naive-product} also offers an
appropriate ansatz for dressing more educated QCD-based predictions
with our EW corrections.  Specifically, replacing
$\sigma^{\mr{NLO\,QCD}}$ by $\sigma^{\mr{NNLO\,QCD}}$, as worked out
in \citere{Grazzini:2013bna}, would deliver state-of-the-art
predictions based on fixed perturbative orders.

\subsection{Virtual corrections}
\label{se:virt}

We calculate the virtual QCD and EW corrections to the partonic processes defined in 
Eqs.~\refeq{eq:LO_zgamma_ll} and \refeq{eq:LO_zgamma_nn}. The QCD corrections include 
contributions from self-energy, vertex, and box (4-point) diagrams only. The virtual EW corrections 
additionally involve pentagon diagrams. The structural diagrams for the EW NLO corrections for process 
\refeq{eq:LO_zgamma_ll} are given in
\reffis{fi:selfenergies}--\ref{fi:pentagons}, and the
pentagons are shown explicitly in \reffi{fi:explicitpentagons}. 
Since the contributions from the LO 
photon--photon-induced contributions are tiny, we neglect EW corrections to this process. 

\begin{figure}
  \begin{center}
  \tabcolsep 5pt
    \begin{tabular}{ccccc} 
      \input{plots/feynman_diagrams/virt/self05.tex} &
      \input{plots/feynman_diagrams/virt/self06.tex} &
      \input{plots/feynman_diagrams/virt/self07.tex} &
      \input{plots/feynman_diagrams/virt/self08.tex} 
\\
      \input{plots/feynman_diagrams/virt/self01.tex} &
      \input{plots/feynman_diagrams/virt/self02.tex} &
      \input{plots/feynman_diagrams/virt/self03.tex} &
      \input{plots/feynman_diagrams/virt/self04.tex} 
    \end{tabular}
  \end{center}
  \vspace*{-2em}
  \caption{Self-energy corrections to the partonic process $\Pq_{i} \, \Pqbar_{i} 
           \to \Plp \Plm \, \gamma$.}
  \label{fi:selfenergies}
  \begin{center}
  \tabcolsep 5pt
    \begin{tabular}{ccccc}
      \input{plots/feynman_diagrams/virt/tri11.tex} &
      \input{plots/feynman_diagrams/virt/tri08.tex} &
      \input{plots/feynman_diagrams/virt/tri06.tex} &
      \input{plots/feynman_diagrams/virt/tri09.tex} &
      \input{plots/feynman_diagrams/virt/tri10.tex} 
      \\
      \input{plots/feynman_diagrams/virt/tri07.tex} &
      \input{plots/feynman_diagrams/virt/tri05.tex} &
      \input{plots/feynman_diagrams/virt/tri02.tex} &
      \input{plots/feynman_diagrams/virt/tri12.tex} &
      \input{plots/feynman_diagrams/virt/tri03.tex} 
      \\
      &
      \input{plots/feynman_diagrams/virt/tri04.tex} &
      \input{plots/feynman_diagrams/virt/tri13.tex} &
      \input{plots/feynman_diagrams/virt/tri01.tex} 
    \end{tabular}
  \end{center}
  \vspace*{-2em}
  \caption{Vertex corrections to the partonic process $\Pq_{i} \, \Pqbar_{i} 
           \to \Plp \Plm \, \gamma$.}
  \label{fi:triangles}
%
  \begin{center}
  \tabcolsep 5pt
    \begin{tabular}{ccc}
      \input{plots/feynman_diagrams/virt/box02.tex} &
      \input{plots/feynman_diagrams/virt/box03.tex} &
      \input{plots/feynman_diagrams/virt/box04.tex} \\
      \input{plots/feynman_diagrams/virt/box01.tex} &
      \input{plots/feynman_diagrams/virt/box05.tex} &
      \input{plots/feynman_diagrams/virt/box06.tex}
    \end{tabular}
  \end{center}
  \vspace*{-2em}
  \caption{Box corrections to the partonic process $\Pq_{i} \, \Pqbar_{i} 
           \to \Plp \Plm \, \gamma$.}
  \label{fi:boxes}
\end{figure}%
\begin{figure}
  \begin{center}
  \tabcolsep 5pt
    \begin{tabular}{c}
      \input{plots/feynman_diagrams/virt/pent01.tex}
    \end{tabular}
  \end{center}
  \vspace*{-2em}
  \caption{Pentagon corrections for the partonic process $\Pq_{i} \, \Pqbar_{i} 
           \to \Plp \Plm \, \gamma$.}
  \label{fi:pentagons}
%
  \begin{center}
  \tabcolsep 5pt
    \begin{tabular}{cccc}
      \input{plots/feynman_diagrams/virt/exppent01.tex} &
      \input{plots/feynman_diagrams/virt/exppent03.tex} &
      \input{plots/feynman_diagrams/virt/exppent05.tex} &
      \input{plots/feynman_diagrams/virt/exppent07.tex} 
      \\
      \input{plots/feynman_diagrams/virt/exppent02.tex} &
      \input{plots/feynman_diagrams/virt/exppent04.tex} &
      \input{plots/feynman_diagrams/virt/exppent06.tex} 
    \end{tabular}
  \end{center}
  \vspace*{-2em}
  \caption{Explicit pentagon diagrams for the partonic process $\Pq_{i} \, \Pqbar_{i} 
           \to \Plp \Plm \, \gamma$.}
  \label{fi:explicitpentagons}
\end{figure}%

Since the $\Pb\Pbbar$~channel contributes only about $3\%$ to the LO
cross section, we omit the corresponding EW corrections which we
expect to be in the sub per-mille level and therefore negligible.

We have performed two independent loop calculations with two different sets of tools,
both making use of traditional methods based on Feynman diagrams. 
The amplitudes are generated in the 't~Hooft--Feynman gauge 
and algebraically reduced using \texttt{MATHEMATICA} programs,
producing a standard representation in terms of standard matrix elements
containing all spinorial and polarization-dependent objects
and Lorentz-invariant coefficients containing the loop integrals.
While the standard matrix elements are evaluated in terms of 
Weyl--van-der-Waerden spinor products following \citere{Dittmaier:1998nn}, 
the loop integrals are computed with the \texttt{COLLIER} library~\cite{Denner:2014gla},
which is based on the results of \citeres{Denner:2002ii,Denner:2005nn,Denner:2010tr}. 
In one calculation we use 
\texttt{FEYNARTS}~3 \cite{Hahn:2000kx,Hahn:2001rv}, \texttt{FORMCALC} \cite{Hahn:1998yk}, and 
\texttt{POLE} \cite{Accomando:2001fn} for the generation and reduction of the amplitudes,
while the second calculation employs inhouse \texttt{MATHEMATICA} routines starting
from amplitudes generated with \texttt{FEYNARTS}~1~\cite{Kublbeck:1990xc}.

\subsection{Real corrections}
\label{se:real} 

The real EW corrections to the quark--antiquark channels are induced
by the partonic processes
\begin{align}
\label{eq:qqEW_real_zgamma_ll}
\Pq_i \,\Pqbar_i &\to \Plp \Plm \, \gamma \, \gamma \, , \\
\label{eq:qqEW_real_zgamma_nn}
\Pq_i \,\Pqbar_i &\to \Panl \Pnl \, \gamma \, \gamma \, .
\end{align}
The Feynman diagrams for \refeq{eq:qqEW_real_zgamma_ll} are shown in
\reffi{fi:qqEW}. 
\begin{figure}     
  \begin{center}
  \tabcolsep 5pt
    \begin{tabular}{ccccc}
      \input{plots/feynman_diagrams/qqEW_real/qqEW_real07.tex} &
      \input{plots/feynman_diagrams/qqEW_real/qqEW_real08.tex} &
      \input{plots/feynman_diagrams/qqEW_real/qqEW_real11.tex} &
      \input{plots/feynman_diagrams/qqEW_real/qqEW_real12.tex} &
      \input{plots/feynman_diagrams/qqEW_real/qqEW_real10.tex} 
\\
      \input{plots/feynman_diagrams/qqEW_real/qqEW_real09.tex} &
      \input{plots/feynman_diagrams/qqEW_real/qqEW_real17.tex} &
      \input{plots/feynman_diagrams/qqEW_real/qqEW_real18.tex} &
      \input{plots/feynman_diagrams/qqEW_real/qqEW_real19.tex} &
      \input{plots/feynman_diagrams/qqEW_real/qqEW_real20.tex} 
\\
      \input{plots/feynman_diagrams/qqEW_real/qqEW_real13.tex} &
      \input{plots/feynman_diagrams/qqEW_real/qqEW_real15.tex} &
      \input{plots/feynman_diagrams/qqEW_real/qqEW_real14.tex} &
      \input{plots/feynman_diagrams/qqEW_real/qqEW_real16.tex} &
      \input{plots/feynman_diagrams/qqEW_real/qqEW_real01.tex} 
      \\
      \input{plots/feynman_diagrams/qqEW_real/qqEW_real02.tex} &
      \input{plots/feynman_diagrams/qqEW_real/qqEW_real06.tex} &
      \input{plots/feynman_diagrams/qqEW_real/qqEW_real03.tex} &
      \input{plots/feynman_diagrams/qqEW_real/qqEW_real05.tex} &
      \input{plots/feynman_diagrams/qqEW_real/qqEW_real04.tex} 
    \end{tabular}
  \end{center}
  \vspace*{-2em}
  \caption{Feynman diagrams of the quark--antiquark-induced real EW corrections 
           for the partonic process $\Pq_{i} \, \Pqbar_{i} \to \Plp \Plm \, \gamma$.}
  \label{fi:qqEW}
\end{figure}
While the production of charged leptons in
\refeq{eq:qqEW_real_zgamma_ll} involves photon emission both from the
IS and FS, the photons in the neutrino production process
\refeq{eq:qqEW_real_zgamma_nn} entirely results from IS radiation
(corresponding to the first six diagrams in
\reffi{fi:qqEW}).  In both processes photon
bremsstrahlung gives rise to soft and collinear singularities when one
of the two photons gets soft or collinear to any charged IS or FS
fermion.  These singularities are extracted from the phase-space
integral and analytically evaluated using the dipole subtraction
technique as formulated for photons in
\citeres{Dittmaier:EWsubtraction,Dittmaier:2008md}.  While the soft
singularities completely cancel against the virtual corrections, the
remaining collinear IS singularities 
can be absorbed into the proton PDFs.
In view of collinear singularities from photon radiation off FS
leptons we have considered 
two scenarios, called the
collinear-safe (CS) and non-collinear-safe (NCS) case in
\citere{Denner:2014bna}.  In detail our calculation closely follows
Section~2.3.1 of \citere{Denner:2014bna}, where the corresponding part
of our NLO calculation for $\PW+\gamma$~production is described.

\begin{figure}
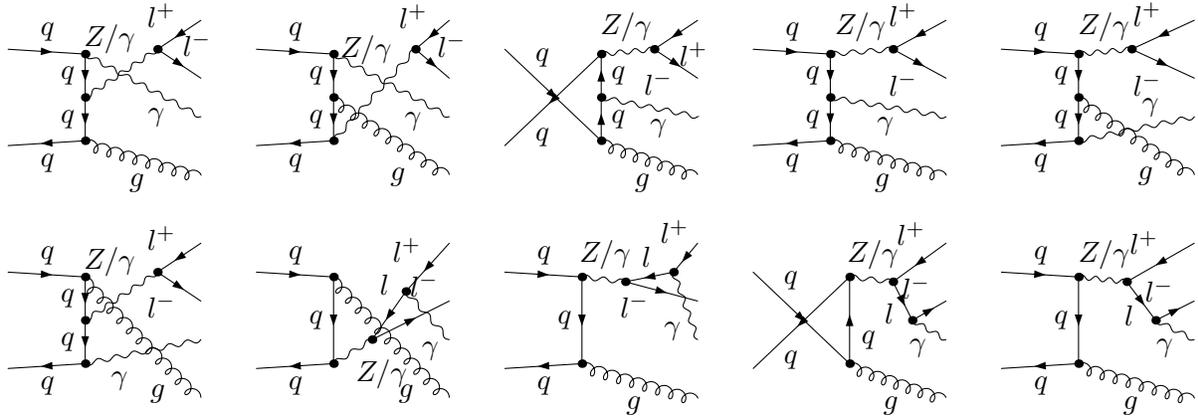
     
  \begin{center}
  \tabcolsep 5pt
    \begin{tabular}{ccccc}
      \input{plots/feynman_diagrams/qqQCD_real/qqQCD_real01.tex} &
      \input{plots/feynman_diagrams/qqQCD_real/qqQCD_real02.tex} &
      \input{plots/feynman_diagrams/qqQCD_real/qqQCD_real05.tex} &
      \input{plots/feynman_diagrams/qqQCD_real/qqQCD_real06.tex} &
      \input{plots/feynman_diagrams/qqQCD_real/qqQCD_real04.tex} 
      \\
      \input{plots/feynman_diagrams/qqQCD_real/qqQCD_real03.tex} &
      \input{plots/feynman_diagrams/qqQCD_real/qqQCD_real07.tex} &
      \input{plots/feynman_diagrams/qqQCD_real/qqQCD_real08.tex} &
      \input{plots/feynman_diagrams/qqQCD_real/qqQCD_real09.tex} &
      \input{plots/feynman_diagrams/qqQCD_real/qqQCD_real10.tex}
    \end{tabular}
  \end{center}
  \vspace*{-2em}
  \caption{Feynman diagrams of the quark--antiquark-induced real QCD
    corrections for the partonic process $\Pq_{i} \, \Pqbar_{i} \to
    \Plp \Plm \, \gamma$.}
  \label{fi:qqQCD}
\end{figure}

For the real QCD corrections we have to consider the partonic channels
\begin{align}
\Pq_i \, \Pqbar_i &\to \Plp \Plm \, \gamma \, \Pg \, , \nonumber \\
\Pq_i \, \Pg      &\to \Plp \Plm \, \gamma \, \Pq_i \, , 
\nonumber \\
\Pqbar_i \, \Pg   &\to \Plp \Plm \, \gamma \, \Pqbar_i \, ,
\label{eq:QCD_real_zgamma_ll}
\end{align}
for the process involving charged leptons and the channels
\begin{align}
\Pq_i \, \Pqbar_i &\to \Panl \Pnl \, \gamma \, \Pg \, , \nonumber \\
\Pq_i \, \Pg      &\to \Panl \Pnl \, \gamma \, \Pq_i \, , 
\nonumber \\
\Pqbar_i \, \Pg   &\to \Panl \Pnl \, \gamma \, \Pqbar_i \, 
\label{eq:QCD_real_zgamma_nn}
\end{align}
for the process with neutrinos in the final state.
The corresponding Feynman diagrams to the first process in \refeq{eq:QCD_real_zgamma_ll} are shown in 
\reffi{fi:qqQCD}. 
The diagrams for the gluon-induced contributions can be derived via crossing symmetries. The 
calculation of these corrections is completely analogous to the one 
described in Sections~2.3.2 and 2.3.3
of \citere{Denner:2014bna}, i.e.\ we again use dipole
subtraction~\cite{Catani:1996vz,Catani:2002hc} to treat soft and
collinear singularities resulting from collinear IS splittings.  Since
the final states in \refeq{eq:QCD_real_zgamma_ll} and
\refeq{eq:QCD_real_zgamma_nn} contain a photon and a jet, which can
become collinear, we apply two different methods for the treatment of
collinear photon--jet configurations. We use the concept of democratic
clustering in combination with a quark-to-photon fragmentation
function as introduced in \citeres{Glover:1993qtp,Glover:1994th} and
the Frixione isolation scheme \cite{Frixione:1998jh}
as an alternative. Both schemes
identify a collinear photon--jet system as a photon if the photonic
energy content in this system exceeds a certain fraction of its total
energy, in order to define the $\PZ+\gamma$ contribution in the
process $\Pp\Pp\to\PZ+\gamma+\mathrm{jet}+X$.  In the same spirit, the
$\PZ+\mathrm{jet}$ contribution was defined in
\citeres{Denner:2011vu,Denner:2012ts}, where NLO QCD+EW corrections to
the complementary processes $\Pp\Pp\to\PZ(\to
l^+l^-/\bar\nu\nu)+\mathrm{jet}+X$ were calculated.

The photon-induced EW corrections include the partonic channels
\begin{align}
\Pq_i \, \gamma      &\to \Plp \Plm \, \gamma \, \Pq_i \, , 
\nonumber \\
\Pqbar_i \, \gamma   &\to \Plp \Plm \, \gamma \, \Pqbar_i \, ,
\label{eq:qph_real_zgamma_ll}
\end{align}
and
\begin{align}
\Pq_i \, \gamma      &\to \Panl \Pnl \, \gamma \, \Pq_i \, , 
\nonumber \\
\Pqbar_i \, \gamma   &\to \Panl \Pnl \, \gamma \, \Pqbar_i \, .
\label{eq:qph_real_zgamma_nn}
\end{align}
The Feynman diagrams for \refeq{eq:qph_real_zgamma_ll} can be derived
from the diagrams in \reffi{fi:qqEW} via crossing of a FS photon to
the IS and a quark or antiquark into the FS. Besides soft and
collinear singularities from photon radiation off fermions and
collinear photon--jet configurations the photon-induced EW corrections
additionally include singularities from the collinear splittings
$\gamma \rightarrow f \bar{f}^{*}$ and $f \rightarrow f \gamma^{*}$.
Our treatment of these singularities follows Sections~3 and 5 of
\citere{Dittmaier:2008md}. Some details can also be found in
Section~2.3.3  of \citere{Denner:2014bna}.

\section{Numerical results}
\label{se:numres}

\subsection{Input parameters and setup}
\label{se:SMinput}

The relevant SM input parameters are
\begin{equation}\arraycolsep 2pt
\begin{array}[b]{rclrclrcl}
\GF & = & 1.1663787 \times 10^{-5} \GeV^{-2}, \quad&
\alpha(0) &=& 1/137.035999074, \quad &
\alpha_{\mathrm{s}}(\MZ) &=& 0.119, 
\\
M_\PH &=& 125\GeV, &
m_\mu &=& 105.6583715\MeV, &
m_\Pt &=& 173.07\GeV,
\\
\MW^{\OS} & = & 80.385\GeV, &
\Gamma_\PW^{\OS} & = & 2.085\GeV, \\
\MZ^{\OS} & = & 91.1876\GeV, &
\Gamma_\PZ^{\OS} & = & 2.4952\GeV. &
\end{array}
\label{eq:SMpar}
\end{equation}
All parameters but $\alpha_{\mathrm{s}}(\MZ)$, which is provided by
the PDF set, are extracted from \citere{Beringer:1900zz}. 
The masses of all quarks but the top quark are set to zero.

Owing to the presence of an on-shell external photon, we always take
one electromagnetic coupling constant $\al$ at zero momentum transfer,
$\al=\al(0)$. For all other couplings, e.g.\ the $\PZ$-boson--fermion
or additional photon--fermion couplings, we determine the
electromagnetic coupling constant in the $G_{\mu}$ scheme, where
$\alpha$ is defined in terms of the Fermi constant,
\begin{align}
   \al_{G_{\mu}}=\frac{\sqrt{2}}{\pi}G_{\mu}M^{2}_{\PW}\klam{1-\frac{M^{2}_{\PW}}{M^{2}_{\PZ}}}.
\end{align}
This definition effectively absorbs some universal corrections into
the LO contributions, such as those associated with the
evolution of $\al$ from zero momentum transfer to the electroweak
scale and corrections related to the $\rho$-parameter.  In
this scheme large fermion-mass logarithms are effectively resummed
leading to an independence of logarithms of the light fermion
masses \cite{Denner:2000bj}
(see also the discussion in the ``EW dictionary'' in \citere{Butterworth:2014efa}).
Using this mixed scheme the squared LO
amplitude is proportional to $\alpha(0)\alpha^{2}_{G_{\mu}}$.
In the relative EW corrections we set the additional coupling factor $\alpha$
to $\al_{G_{\mu}}$, because this coupling is adequate for the most pronounced
EW corrections which are caused by soft/collinear weak gauge-boson exchange
at high energies (EW Sudakov logarithms, etc.).

We use the complex-mass scheme
\cite{Denner:1999gp,Denner:2005fg,Denner:2006ic} to treat the Z-boson
resonance by introducing complex vector-boson masses $\mu_{\PW,\PZ}$
according to
\begin{align}
M^{2}_{\PW}  \rightarrow \mu^{2}_{\PW} = M^{2}_{\PW}-\ri M_{\PW}\Gamma_{\PW}\,, \qquad
M^{2}_{\PZ}  \rightarrow \mu^{2}_{\PZ} = M^{2}_{\PZ}-\ri M_{\PZ}\Gamma_{\PZ}
\end{align}
with constant decay widths $\Gamma_{\PW,\PZ}$.  However, at LEP and the Tevatron the on-shell
(OS) masses of the vector bosons were measured, which correspond to
running widths. Therefore, the OS masses $M^{\OS}_{\PW}$, $M^{\OS}_{\PZ}$ and
widths $\Gamma^{\OS}_{\PW}$, $\Gamma^{\OS}_{\PZ}$ have to be
converted to the pole values using the relations
\cite{Bardin:1988xt}
\begin{align}
  M_{\mathrm{V}}=M^{\OS}_{\mathrm{V}}/\sqrt{1+\klam{\Gamma^{\OS}_{\mathrm{V}}/M^{\OS}_{\mathrm{V}}}^{2}}, 
     \qquad \Gamma_{\mathrm{V}}=
            \Gamma^{\OS}_{\mathrm{V}}/\sqrt{1+\klam{\Gamma^{\OS}_{\mathrm{V}}/M^{\OS}_{\mathrm{V}}}^{2}}
     \qquad (\mathrm{V}=\mathrm{W},\;\PZ)\,,
\end{align}
resulting in
\begin{align}
  \begin{array}[b]{r@{\,}l@{\qquad}r@{\,}l}
\MW &= 80.3580\ldots\GeV, & \Gamma_{\PW} &= 2.0843\ldots\GeV, \\
\MZ &= 91.1535\ldots\GeV,& \Gamma_{\PZ} &= 2.4943\ldots\GeV.
\label{eq:m_ga_pole_num}
\end{array}
\end{align}

Calculating the hadronic cross section, we employ the NNPDF2.3QED PDF
set~\cite{Ball:2013hta}, which includes a photon PDF, QED
contributions to parton evolution and the two-loop running of
$\al_{s}$ for five active flavours ($n_{f}=5$). Following the
arguments of \citere{Diener:2005me}, we apply a DIS-like factorization
scheme for the QED corrections \change{(see, e.g.,
  \citere{Dittmaier:2009cr})}, but an $\overline{\mathrm{MS}}$
prescription for the QCD corrections as demanded by the NNPDF2.3QED
PDF set.

\change{Strictly speaking, the choice of the factorization scheme of
  the QED corrections is ambiguous for the NNPDF2.3QED PDF set.
  Therefore, we have performed the calculation for the QED corrections
  also using the $\overline{\mathrm{MS}}$ factorization scheme.  The
  results for the integrated cross sections differ from those obtained
  with the DIS-like scheme by less than $0.05\%$ both for the relative
  EW corrections and the photon-induced corrections
  relative to the leading order. Also in all considered distributions
  the changes are well below $0.1\%$ and thus phenomenologically
  negligible.}

The factorization and the renormalization scales $\mu_{\mathrm{F}}, \mu_{\mathrm{R}}$
are set equal throughout our calculation.
Following \citeres{Dixon:1999di,Haywood:1999qg}, we choose the scales as
\begin{align}
  \mu^{2}_{\mathrm{F}}=\mu^{2}_{\mathrm{R}}=\frac{1}{2}\klam{\MZ^{2}+p^{2}_{\mathrm{T},\PZ}
          +p_{\mathrm{T},\gamma_{1}}^{2}+p_{\mathrm{T},\gamma_{2} / \mathrm{jet}}^{2}},
\end{align}
where $p_{\mathrm{T},\PZ}$ is the transverse momentum of the massive vector boson defined by
\begin{align}
  p_{\mathrm{T},\PZ} = \left\{
\begin{array}{l l}%
|{\bf p}_{\mathrm{T},\Panl}+{\bf p}_{\mathrm{T},\Pnl}|\qquad 
&{\rm for\quad }\Pp \Pp  \rightarrow \bar{\nu}_l \nu_l \gamma,\\[1ex]
|{\bf p}_{\mathrm{T},\Plp}+{\bf p}_{\mathrm{T},\Plm}|\qquad 
&{\rm for\quad }\Pp \Pp \rightarrow l^+l^- \gamma,\\
 \end{array}
\right.
\label{eq:pt_boson}
\end{align}
and $p_{\mathrm{T},a}=|{\bf p}_{\mathrm{T},a}|$ denotes the absolute
value of the transverse three-momentum ${\bf p}_{\mathrm{T},a}$ of
particle $a$.  The photons $\ga_1$ and $\ga_2$ are ordered so that
$p_{\mathrm{T},\gamma_{1}}>p_{\mathrm{T},\gamma_{2}}$ and we call
the hardest photon the one with the highest transverse momentum. 
In LO the transverse momenta $p_{\mathrm{T},\gamma_{2} /
  \mathrm{jet}}$ vanish.

The QCD scale uncertainty of $\PZ+\gamma$ production has already been
investigated in various publications such as in
\citeres{Ohnemus:1994qp,DeFlorian:2000sg,Campbell:2011bn}.  Varying
the scale by a factor of two the scale dependence was found to be of
the order of $5\%$ at NLO QCD in \citere{Campbell:2011bn}.  Meanwhile
NNLO QCD corrections have been calculated~\cite{Grazzini:2013bna} and
found to contribute another $6\%$ on top of the NLO QCD prediction for
the integrated cross section.  The corresponding scale uncertainty is
reduced to $2\%$ at NNLO QCD. Note, however, that the scale definition
slightly differs from ours.

\subsection{Phase-space cuts and event selection}
\label{se:cuts}
The processes $\Pp\Pp\to \Panl \Pnl + \gamma + X$ and $\Pp\Pp\to \Pl^+
\Pl^- + \gamma + X$ require the recombination of FS photons with FS
partons and, in case of the second process, of FS photons with charged
leptons in regimes of phase space where photon and parton/lepton are
collinear.  Furthermore, we impose several cuts to account for the
detector acceptance. The phase-space cuts and the event selection are
inspired by the recent ATLAS and CMS papers
\cite{Aad:2013izg,Chatrchyan:2013fya,Aad:2014fha,Khachatryan:2015kea}
analyzing $\PW\gamma$ and $\PZ\gamma$ final states.

\subsubsection{Recombination}
Recombination of a photon and a FS particle is based on 
the Euclidean distance in the $y$--$\phi$ plane,
$R_{ij}=\sqrt{\klam{y_{i}-y_{j}}^{2}+\phi^{2}_{ij}}$, where 
$y=\frac{1}{2}\ln\left[\klam{E+p_{\mathrm{L}}}/\klam{E-p_{\mathrm{L}}}\right]$
denotes the rapidity.
Here, $E$ is the energy and $p_{\mathrm{L}}$ the
longitudinal momentum of the respective particle along the
beam axis. Furthermore,  $\phi_{ij}$ refers to the angle between the
particles $i$ and $j$ in the plane perpendicular to the beams.  The
recombination is performed as follows:
\begin{enumerate}
\item If we consider ``bare'' muons, a photon and a charged
  (anti)lepton are never recombined.  Otherwise recombination is
  applied if $R_{l^{\pm}\gamma}<0.1$, and the four-momenta of photon
  and lepton are added.  If the separation in $R$ between the photon
  and each of the two leptons is smaller than $0.1$ at the same time,
  the photon is recombined with the lepton that has a smaller
  $R_{l\gamma}$ separation.  In case of two photons in the final
  state, first the photon with the smaller $R_{l^{\pm}\gamma}$ is
  recombined.
\item Two photons are recombined if $R_{\gamma\gamma}<0.1$.
\item 
  Using the method of democratic clustering,
  a photon and a jet are recombined if their distance
  becomes $R_{\gamma\mathrm{jet}}<R_0=0.5$. 
  After recombination, the energy fraction
  $z_{\gamma}=E_{\gamma}/\klam{E_{\gamma}+E_{\mathrm{jet}}}$ of the
  photon inside the photon--jet system
  is determined.
  If $z_{\gamma}$ is smaller than $z_{\mr{cut}}=0.9$ the
  event is regarded as a part of the process $\PZ+\mathrm{jet}$ and 
  therefore rejected.
\end{enumerate}
The case where more than two particles are recombined is excluded by
our basic cuts.  Results are presented for ``bare'' muons and for
photon recombination with leptons. The latter results 
hold for electrons as well as for muons, since the
lepton-mass logarithms cancel as dictated by the KLN theorem
\cite{Kinoshita:1962ur,Lee:1964is}.

If alternatively the Frixione isolation scheme is applied, step~3 is
replaced as follows:
\begin{enumerate}
\item[3'.]
If $R_{\gamma\mathrm{jet}}<R_{0}=0.5$ the photon and the jet are 
recombined and the event is only accepted if it respects the inequality
\begin{align}
\label{eq:frixione}
  p_{\mr{T},\mr{jet}} < \varepsilon \, p_{\mr{T},\gamma} 
   \klam{\frac{1-\cos \klam{R_{\ga\mr{jet}}}}{1-\cos \klam{R_{0}}}}.
\end{align}
This condition replaces  the condition $z_{\gamma} > z_{\mr{cut}}$ used in the
approach based on democratic clustering and the quark-to-photon 
fragmentation function. Neglecting the difference between $E$ and
$p_{\mr{T}}$ and taking into account that $R_{\gamma \mr{jet}}\sim R_{0}$
for the critical events,
the two parameters $ z_{\mr{cut}}$ and $\varepsilon$ can be related by
\begin{align} \label{eq:zcut_eps_relation}
  z_{\mr{cut}}\approx\frac{1}{1+\varepsilon}.
\end{align}
With this equation we 
get $\varepsilon=0.11$ for $z_{\mr{cut}}=0.9$. 
\end{enumerate}

\subsubsection{Basic cuts}
\label{sec:basic-cuts}
After recombination, we define events for $\Pp\Pp\to \Pl^+ \Pl^- + \gamma + X$ 
by the following cut procedure:
\begin{enumerate}
\item We demand two charged leptons with transverse momentum 
  $p_{\mathrm{T},l^{\pm}}>25 \;\mathrm{GeV}$.
\item We require at least one photon with transverse momentum
  $p_{\mathrm{T},\gamma}>15 \;\mathrm{GeV}$ that is isolated from the
  charged leptons with a distance $R_{l^{\pm}\gamma}>0.7$.
\item The charged leptons and the hardest photon passing the cuts at
  step 2 have to be central, i.e.\ their rapidities have to be in the
  range $\abs{y}<2.5$.
\item 
Only events with an invariant mass of the lepton pair 
$M_{\Plp\Plm}>40\;\GeV$ are accepted, where
\begin{align}
\label{eq:Mllinv}
 M_{\Plp\Plm} & =  \sqrt{\left(p_{\Plp}+p_{\Plm}\right)^{2}} ,
\end{align} 
and $p_{\Plp}$ and $p_{\Plm}$ are the four-vectors of the charged leptons. 
\end{enumerate}

Events for the process $\Pp\Pp\to \Panl \Pnl + \gamma + X$ are defined
by the following cut procedure:
\begin{enumerate}
\item We demand a missing
  transverse momentum $\slashed{p}_{\mathrm{T}}>90 \;\mathrm{GeV}$,
  where $\slashed{p}_{\mr{T}}$ is given by
  \begin{align}
  \label{eq:pt_miss}
   \slashed{p}_{\mr{T}}= |{\bf p}_{\mathrm{T},\Panl}+{\bf p}_{\mathrm{T},\Pnl}|.
  \end{align}
\item We require at least one photon with transverse momentum
  $p_{\mathrm{T},\gamma}>100 \;\mathrm{GeV}$.
\item The hardest photon passing the cuts at step 2 has to be central,
  i.e.\ its rapidity has to be in the range $\abs{y_\gamma}<2.5$.
\item Only events with $\phi_{\gamma,\,\mr{miss}}>2.6$ are taken into
  account, where $\phi_{\gamma,\,\mr{miss}}$ is the angle between the
  missing transverse momentum $\slashed{\bf p}_{\mr{T}}= {\bf
    p}_{\mr{T},\Panl}+{\bf p}_{\mr{T},\Pnl}$ and the hardest photon
  momentum in the plane perpendicular to the beam axis.
\end{enumerate}

We present results with and without applying a jet veto. 
Applying a jet veto means
that all events including a FS jet with $p_{\mr{T,\mathrm{jet}}}>100\GeV$
are discarded. 
Experimentally a jet is required to lie in the rapidity range
$\abs{y_{\mathrm{jet}}}<4.4$. In our calculation we do not restrict the
rapidity range of the vetoed jets, since the related impact on the cross
section is very small and lies within the theoretical uncertainty.

\subsection{\boldmath{Dilepton + photon production: $\Pp\Pp\to \Pl^+ \Pl^- + \gamma + X$}}
\label{se:zgamma_ll}

\subsubsection{Results on total cross sections}
\label{se:CSresults_zgamma_ll}

In \refta{ta:totcs_Zll} we present the LO cross sections
$\si^{\mr{LO}}$ for different $\Pp\Pp$ centre-of-mass energies
$\sqrt{s}$ and different types of relative corrections $\de$ defined
in \refeq{eq:relcor} for $\Pp\Pp\to \Pl^+ \Pl^- + \gamma + X$.
\begin{table}
  \centering\renewcommand{\arraystretch}{1.2}
  \begin{tabular}[H]{c|rrr}
    \multicolumn{4}{c}{$\Pp\Pp \to \Plp \Plm \gamma + X $} \\
    \hline\hline
    $\sqrt{s}/\TeV\; $ & $7$\phz & $8$\phz & $14$\phz \\
    \hline\hline
    $\si^{\mr{LO}}             /\fba\;$ &$\;728.85(4)$&$\;818.43(5)$&$\;1317.4(1)$  \\ 
    \hline \hline
    $\de_{\phot, \Pq\overline{\Pq}}^{\mr{NCS}}/\% \;$   
    &$\; -4.79(2)$ &$\; -4.76(2)$ &  $\; -4.70(2)$     \\
    \hline 
    $\de_{\phot, \Pq\overline{\Pq}}^{\mathrm{CS}}/\% \;$    
    &$\; -2.74(1)$ &$\; -2.73(1)$ & $\; -2.70(1)$      \\ 
    \hline 
    $\de_{\mr{weak}, \Pq\overline{\Pq}}/\% \;$    
    &$\; -0.73$\phz &$\; -0.73$\phz & $\; -0.74$\phz      \\ 
    \hline \hline
    $\de^{\mr{frag}}_{\EW, \Pq\ga}/\% \;$        &$\; 0.04$\phz &$\; 0.04$\phz &$\; 0.04$\phz         \\ 
    \hline 
    $\de^{\mr{veto},\,\mr{frag}}_{\EW, \Pq\ga}/\% \;$ &$\; 0.02$\phz &$\; 0.02$\phz &$\; 0.02$\phz         \\ 
    \hline
    $\de^{\mr{Frix}}_{\EW, \Pq\ga}/\% \;$        &$\; 0.04$\phz &$\; 0.04$\phz &$\; 0.05$\phz         \\ 
    \hline 
    $\de^{\mr{veto},\,\mr{Frix}}_{\EW, \Pq\ga}/\% \;$ &$\; 0.02$\phz &$\; 0.02$\phz &$\; 0.02$\phz         \\ 
    \hline 
    $\de_{\gamma\gamma}/\% \;$ &$\; 0.27$\phz &$\; 0.26$\phz &$\; 0.22$\phz         \\ 
    \hline \hline
    $\de^{\mr{frag}}_{\QCD}/\% \;$        &$\; 61.48(5)$ &$\; 62.90(5)$ &$\; 67.58(5)$    \\ 
    \hline
    $\de^{\mr{Frix}}_{\QCD}/\% \;$        &$\; 60.62(4)$ &$\; 61.96(5)$ &$\; 67.09(7)$    \\ 
    \hline
    $\de^{\veto,\,\mr{frag}}_{\QCD}/\% \;$           &$\; 58.76(5)$ &$\; 59.69(5)$ &$\; 63.11(6)$     \\ 
    \hline
    $\de^{\veto,\,\mr{Frix}}_{\QCD}/\% \;$           &$\; 57.76(4)$ &$\; 58.86(6)$ &$\; 62.33(5)$     \\ 
    \hline \hline
  \end{tabular}
  \caption{\label{ta:totcs_Zll} 
    Integrated cross sections  and relative corrections for $\Pp\Pp
    \to \Plp \Plm \gamma + X $  
    at different LHC energies. The EW corrections to the quark--antiquark 
    annihilation channels are split into purely weak and photonic
    corrections. The photonic corrections  
    are provided with (CS) and  without (NCS)
    lepton--photon recombination. Contributions from the
    photon-induced channels and QCD corrections are shown with a jet
    veto (veto) as well as without a jet veto,
    using a fragmentation function (frag) or
    the Frixione isolation criterion (Frix) to separate photons and jets.
    The numbers in
    parentheses denote the integration errors in the last digits. This
    error is omitted if it is negligible at the given accuracy.} 
\end{table}
As already mentioned in Sect.~\ref{se:setup}, we split the EW
corrections according to Eq.~\refeq{eq:delta_qed_weak} into the
photonic and the weak contributions $\delta_{\phot}$ and
$\delta_{\mr{weak}}$, respectively.  For the photonic corrections
resulting from the quark--antiquark-induced channels we show results
for the CS and NCS scenarios.  Results for the EW corrections
originating from photon-induced channels and for the QCD corrections
are listed with and without a jet veto.  Furthermore, we present
results obtained by applying democratic clustering in combination with
a quark-to-photon fragmentation function and the Frixione isolation
scheme indicated by ``frag" and ``Frix", respectively.  The different
relative corrections are not particularly sensitive to the collider
energy. The largest variation ($\sim60{-}68\%$) occurs in the QCD
corrections.  A jet veto allowing a maximal jet transverse momentum of
$100\GeV$ does not diminish the QCD corrections considerably, since
energy scales dominating the integrated cross section are much lower
for our setup, which allows for photons (leptons) down to
transverse-momentum values of $15\,(25)\GeV$.  The gluon-induced
channels (not separately shown) contribute only about a tenth to the
QCD corrections at an energy of $14\TeV$ and even less at lower
collider energies.  The results obtained with the fragmentation
function and the Frixione isolation scheme differ by $0.5{-}1\%$ for
the QCD corrections.  The photonic corrections to the quark--antiquark
channels are about $-2.7\%$ and $-4.7\%$ for the CS and the NCS case,
respectively.  The weak corrections are about $-0.7\%$ almost
independent of the collider energy.  The quark--photon-induced
corrections contribute less than $0.05\%$ with and without a jet veto
and, thus, are phenomenologically negligible.  The
photon--photon-induced channel contributes with $\sim 0.25\%$.

In summary, the quark--antiquark-induced EW corrections to the
integrated cross sections are small compared to the NLO QCD
corrections. Nevertheless, in particular, the photonic corrections
become relevant in future analyses, since
they are of the order of several percent, i.e.  larger than the
residual scale uncertainty of the NNLO QCD corrections. The
photon-induced EW corrections are at the per-mille level and not
significant for experimental cross-section measurements.  However,
larger effects appear in differential distributions, as
demonstrated in the following.

\subsubsection{Results on transverse-momentum distributions}
\label{se:results_mom_distr_zgamma_ll}

In the following we present differential distributions including QCD
and EW corrections to $\Pp\Pp\to \Pl^+ \Pl^- + \gamma + X$ for a
$\Pp\Pp$ centre-of-mass energy of $14 \TeV$.  For each distribution
the relative EW corrections of the $\Pq\overline{\Pq}$, $\Pq\gamma$,
and $\gamma\gamma$ channels as well as the QCD corrections with and
without a jet veto are shown.  Since the difference between Frixione
isolation and the quark-to-photon fragmentation function is of the
order of $1\%$ for the integrated cross section and distributions, and
therefore not very significant, we only show results obtained with the
quark-to-photon fragmentation function.  For $\PZ+\gamma$ production
the purely weak and the photonic corrections can be separated in a
gauge-independent way. In order to show the impact of the weak
corrections $\delta_{\mr{weak},\, \Pq\overline{\Pq}}$ we plot them
additionally to the full EW corrections.

\begin{figure}     
 \centerline{
         \includegraphics[scale=0.6]{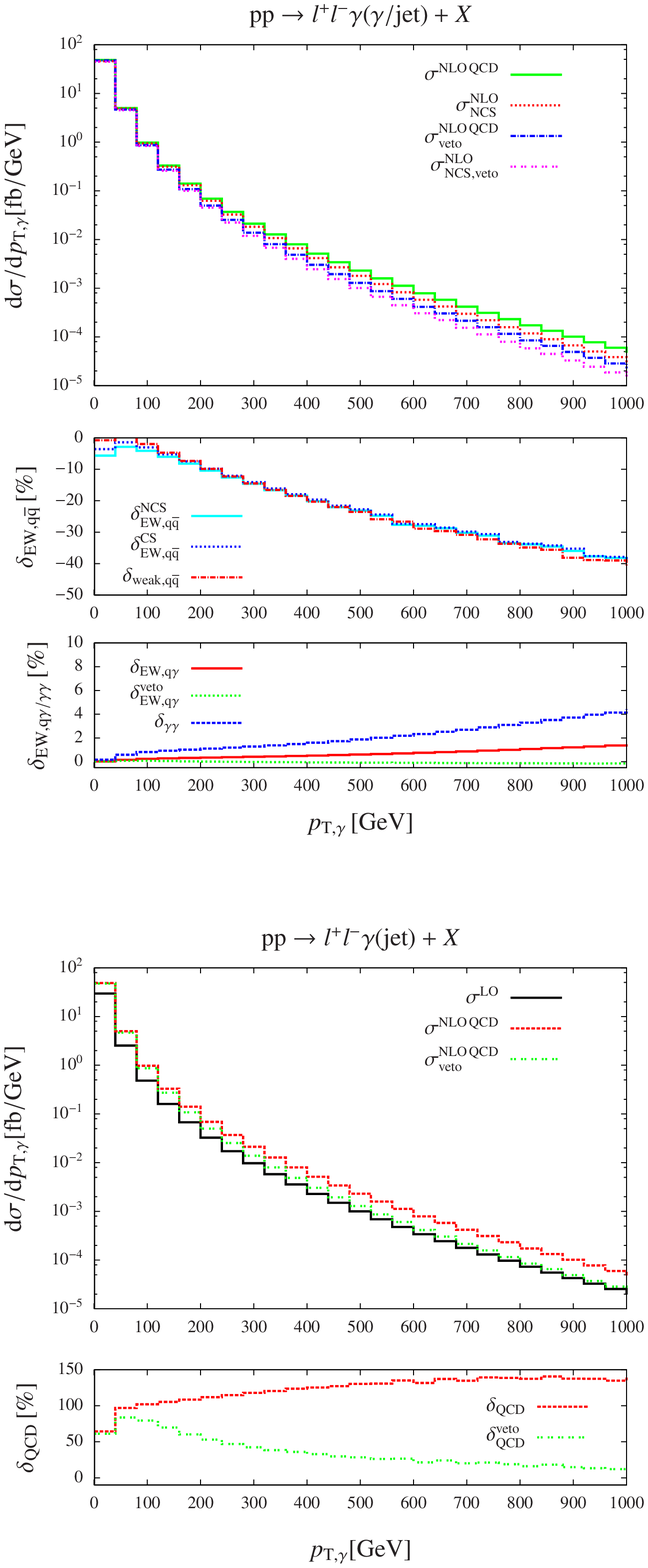}
 \hspace{-4em}
         \includegraphics[scale=0.6]{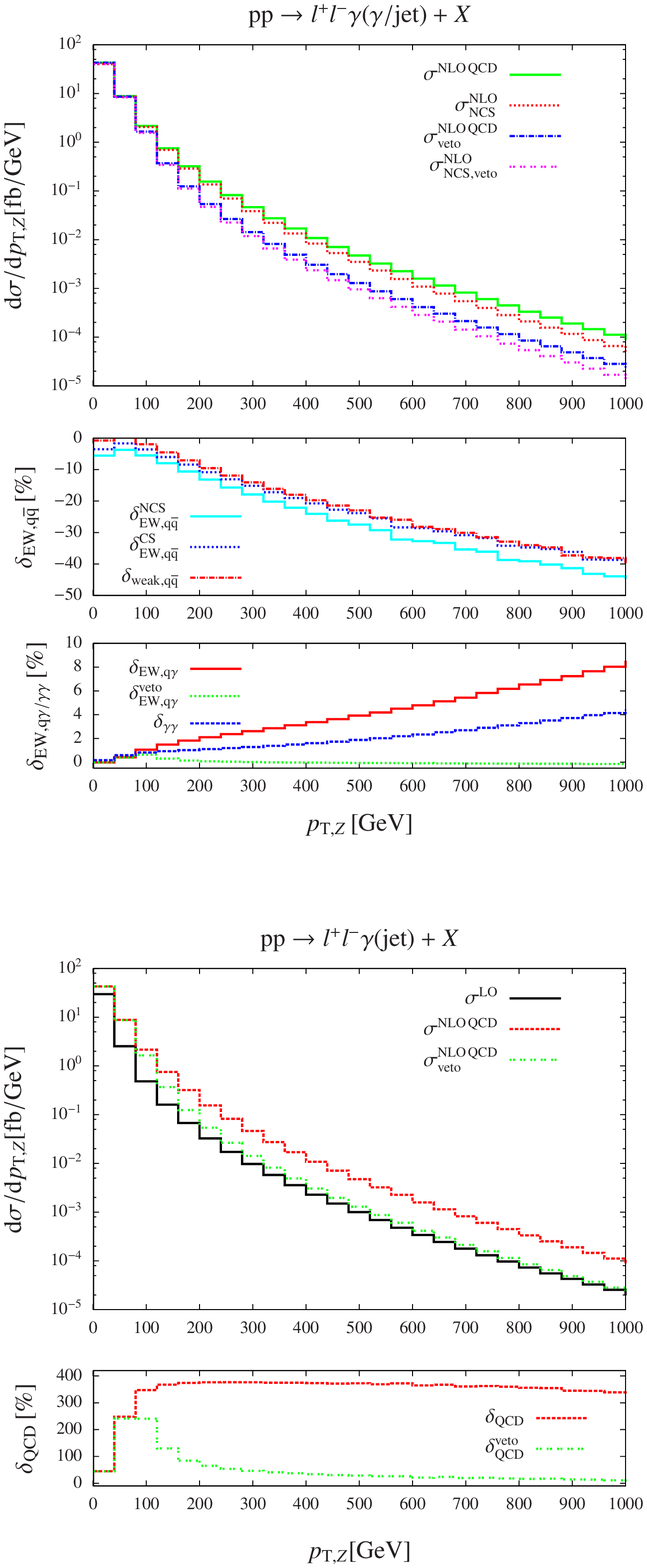}
 }
\caption{\label{fi:pt_boson_zll} Distributions in the
  transverse momentum $p_{\mathrm{T}}$ of the hardest photon (left)
  and the Z~boson (right), including EW (top) and QCD corrections
  (bottom). The large boxes show absolute predictions, the small ones
  relative corrections.}
\end{figure}

In \reffi{fi:pt_boson_zll} we show results on the 
transverse-momentum distributions of the hardest photon (within cuts) 
and of the $\PZ$~boson (defined in Eq.~\refeq{eq:pt_boson}).
Both distributions receive large QCD corrections in the region of high
transverse momenta. This is due to the fact that at NLO QCD new
channels appear ($\Pq g \rightarrow \Plp\Plm\gamma\Pq$) causing large
corrections, especially in the high-$p_{\mr{T}}$ tails.  However,
these large corrections originate from events with hard jets. These
events should preferably be considered as part of $\PZ+\mr{jet}$
rather than $\PZ+\gamma$ production.  Therefore we additionally show
distributions for the case of a jet veto discarding events with
$p_{\mr{T,\,jet}}>100\GeV$. The jet veto suppresses the large QCD
corrections at high transverse momenta.  The $p_{\mr{T}}$
distributions of the photon and the $\PZ$ boson receive large negative
EW corrections, which predominantly originate from so-called EW
Sudakov logarithms included in the weak corrections
$\delta_{\mr{weak},\, \Pq\overline{\Pq}}$. In case of the
$p_{\mr{T},\gamma}$ distribution the CS and the NCS cases hardly
differ, since the recombination of the second photon and a collinear
lepton hardly influences the transverse momentum of the hardest
photon. By contrast, the CS and the NCS cases differ in the
$p_{\mr{T},\PZ}$ distribution.  This is due to the fact that the
transverse momentum of the $\PZ$~boson is reconstructed from the
momenta of the charged leptons which are sensitive to the
recombination with a collinearly radiated photon. The
quark--photon-induced corrections are below $10\%$ in both
distributions and almost vanish in case of a jet veto. The
photon--photon-induced corrections grow up to $4\%$ at
$p_{\mr{T},\PZ}=1\TeV$.  They are not affected by the jet veto, since
there is no jet in the FS.  In summary, the EW corrections are much
smaller than the QCD corrections if no jet veto is applied, but
sizeable.  In case of a jet veto they even become the leading
corrections in the high-transverse-momentum tails.
 
\begin{figure}     
 \centerline{
         \includegraphics[scale=0.6]{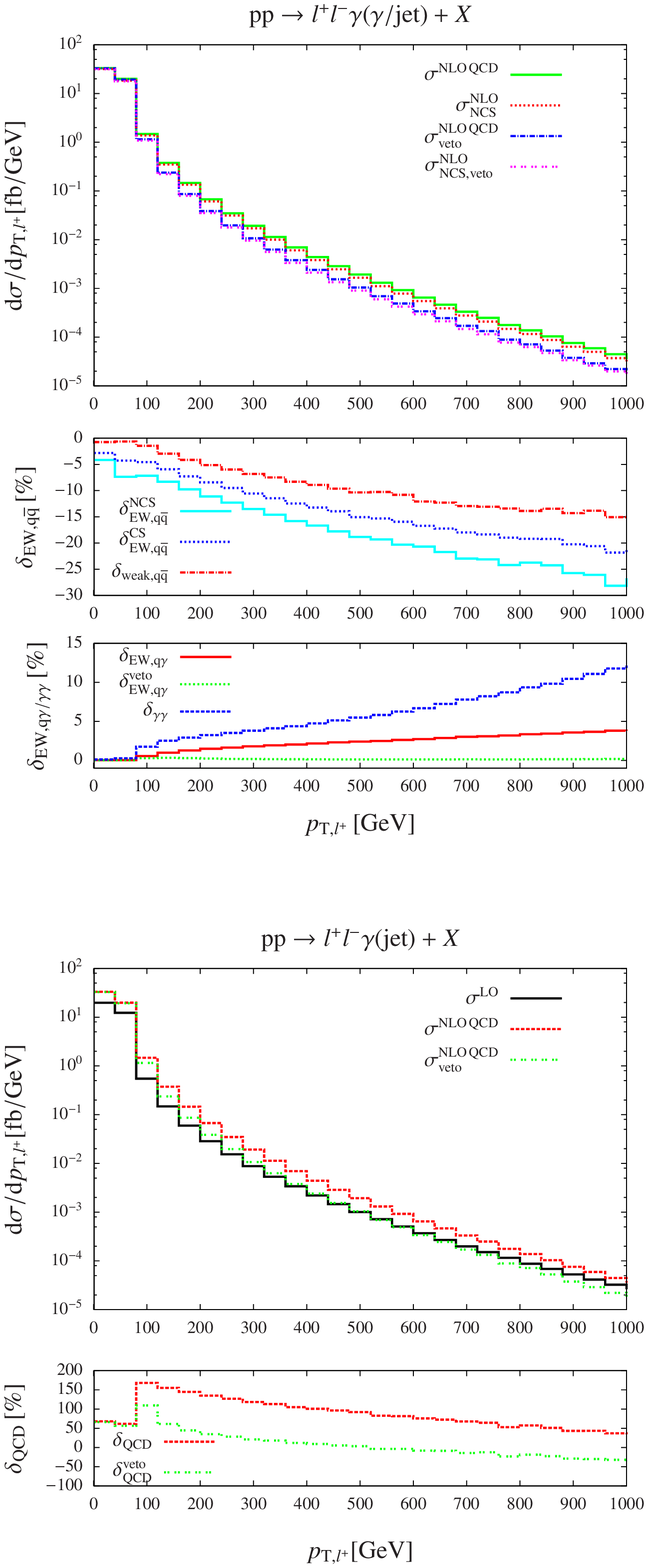}
 \hspace{-4em}
         \includegraphics[scale=0.6]{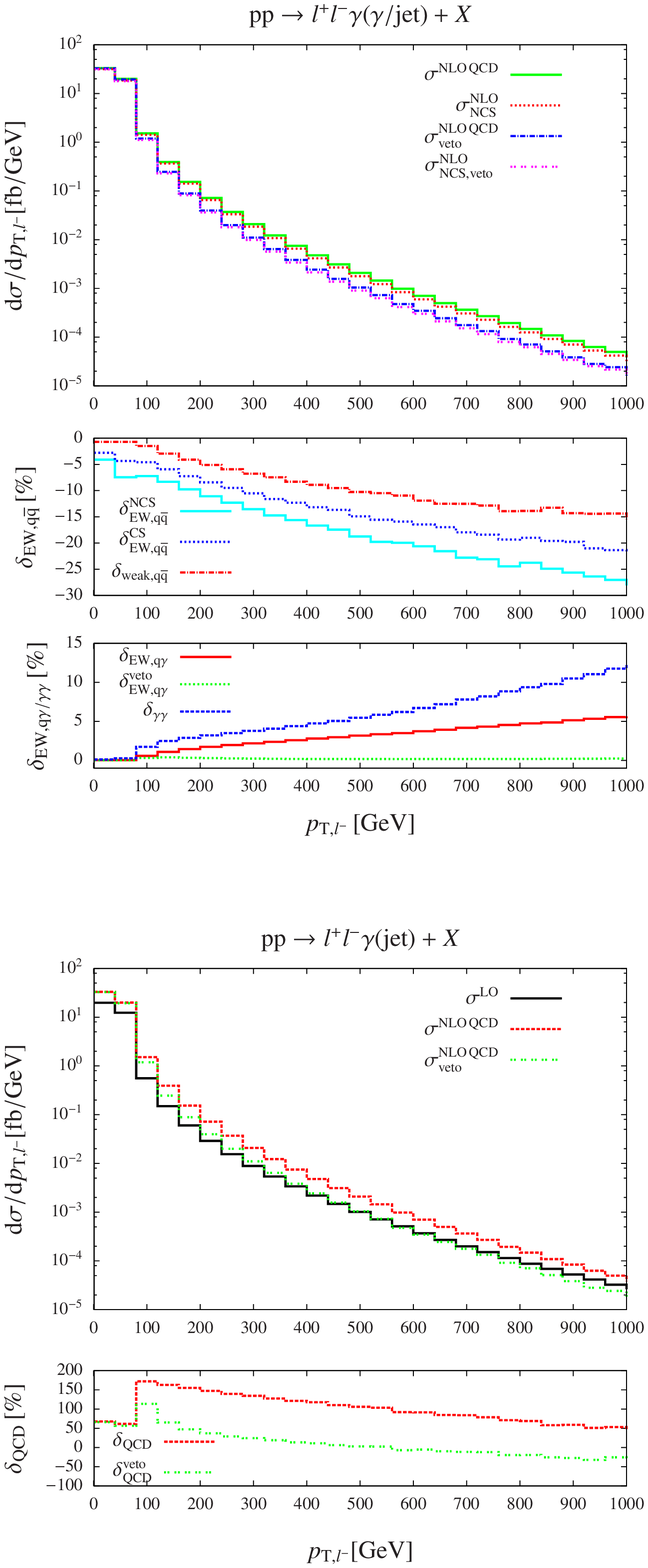}
 }
\caption{\label{fi:pt_lepton_zll} Distributions in the
  transverse momenta $p_{\mathrm{T},l^\pm}$ of the two charged leptons, 
   including EW (top) and QCD corrections (bottom). The large boxes show
  absolute predictions, the small ones relative corrections.}
\end{figure}

The transverse-momentum distributions of the two charged leptons are shown in \reffi{fi:pt_lepton_zll}.
The QCD corrections turn out to be of the order of $150\%$ at
$100\GeV$ and decrease to $50\%$ at $1 \TeV$ if no jet veto is
applied. In case of a jet veto the corrections are still large
($100\%$) in the low $p_{\mr{T}}$-range and drop to $-50\%$ at $1
\TeV$.  The transverse-momentum distribution of each charged lepton
receives large negative weak corrections originating from the Sudakov
logarithms, reaching $-15\%$ at $1 \TeV$. The difference between the
CS and the NCS EW corrections is roughly $6\%$. The collinear
radiation of photons off FS charged leptons shifts the lepton
transverse momentum to smaller values, causing negative corrections.
Recombining the charged lepton with the collinear photon partly
compensates this effect, which is why the CS corrections are smaller.
The quark--photon-induced corrections are below $5\%$ and almost
vanish in case of a jet veto. The photon--photon-induced correction
grows up to more than $10\%$ at $1 \TeV$.  
In the high-$p_{\mr{T}}$ tail the EW corrections
are of the same order of magnitude as the QCD corrections with and
without a jet veto.  The transverse-momentum distributions of the two
charged leptons and the corresponding corrections do not differ
significantly.

\change{The large EW and photon-induced corrections at high transverse
  momenta and invariant masses raise the question of the corresponding
  uncertainties. The leading EW corrections in this region arise from
  the Sudakov double logarithms which are of purely weak origin and known to exponentiate
  \cite{Fadin:1999bq}. Therefore, we can estimate the uncertainty from
  the missing NNLO EW corrections as the square of the relative NLO weak
  corrections $(\delta_{\mr{weak},\, \Pq\overline{\Pq}})^2$, which
  amounts to $16\%$ for the $p_{\mr{T},\gamma}$ $p_{\mr{T},\PZ}$
  distributions and $2\%$ for the distributions in the transverse
  momenta of the leptons at $1\TeV$. This estimate is in agreement
  with calculations of NNLO EW Sudakov corrections for processes with
  on-shell vector bosons \cite{Kuhn:2011mh}.
  
  At large Bjorken-$x$ the photon-PDF carries large uncertainties of
  the order of $100\%$ \cite{Ball:2013hta}. This can be translated to
  an uncertainty for the photon-induced processes where these yield
  large contributions.  Therefore, the contributions of the
  photon-induced processes should be viewed as an uncertainty for our
  predictions. It is negligible where the contributions of
  photon-induced processes are small, but relevant once these get of
  the order of a percent.
  
  The recipes of the previous two paragraphs can also be used for the
  following distributions. However, since the corrections are mostly
  smaller this is also the case for the uncertainties which typically
  can be considered to be at the level of $1\%$ unless the photonic
  corrections exceed $1\%$ or the weak corrections exceed $10\%$.}

\subsubsection{Results on invariant-mass distributions}
\label{se:results_mass_distr_zgamma_ll}

\begin{figure}
 \centerline{
         \includegraphics[scale=0.6]{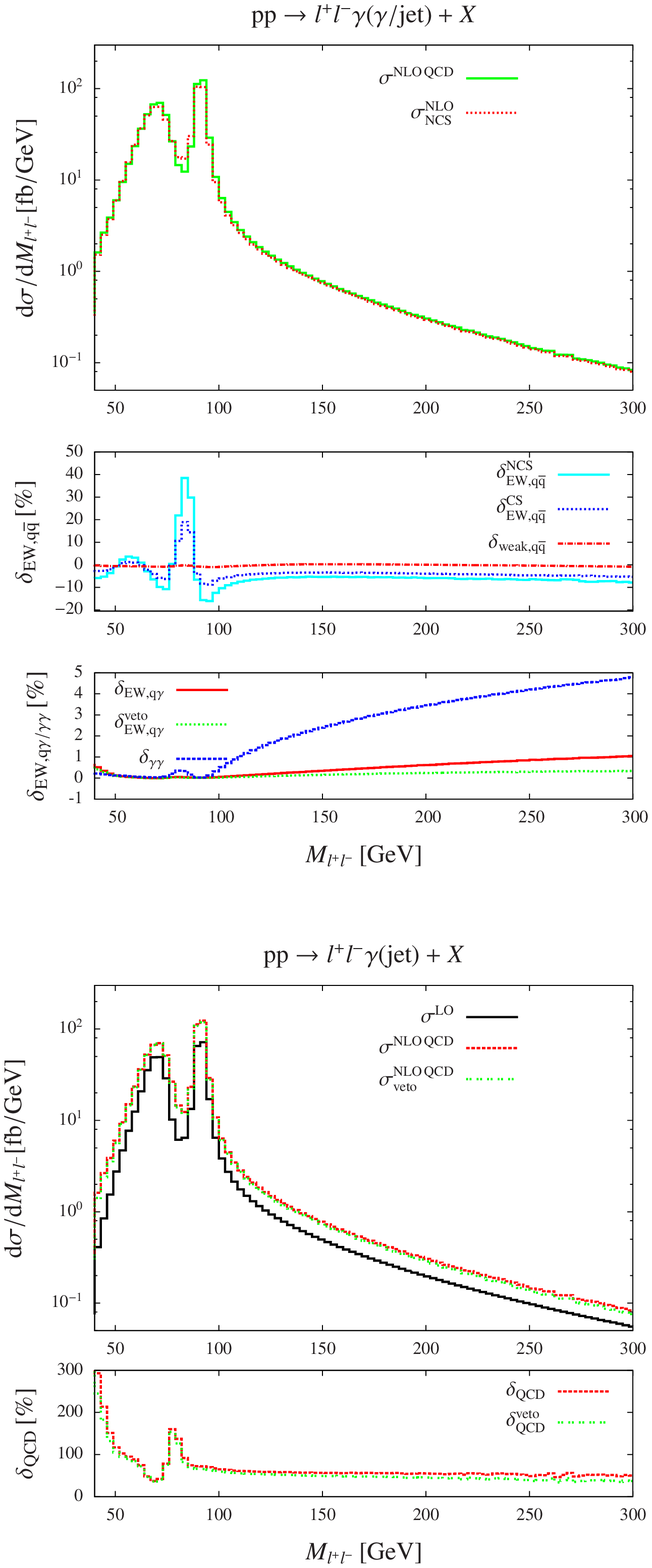}
 \hspace{-4em}
         \includegraphics[scale=0.6]{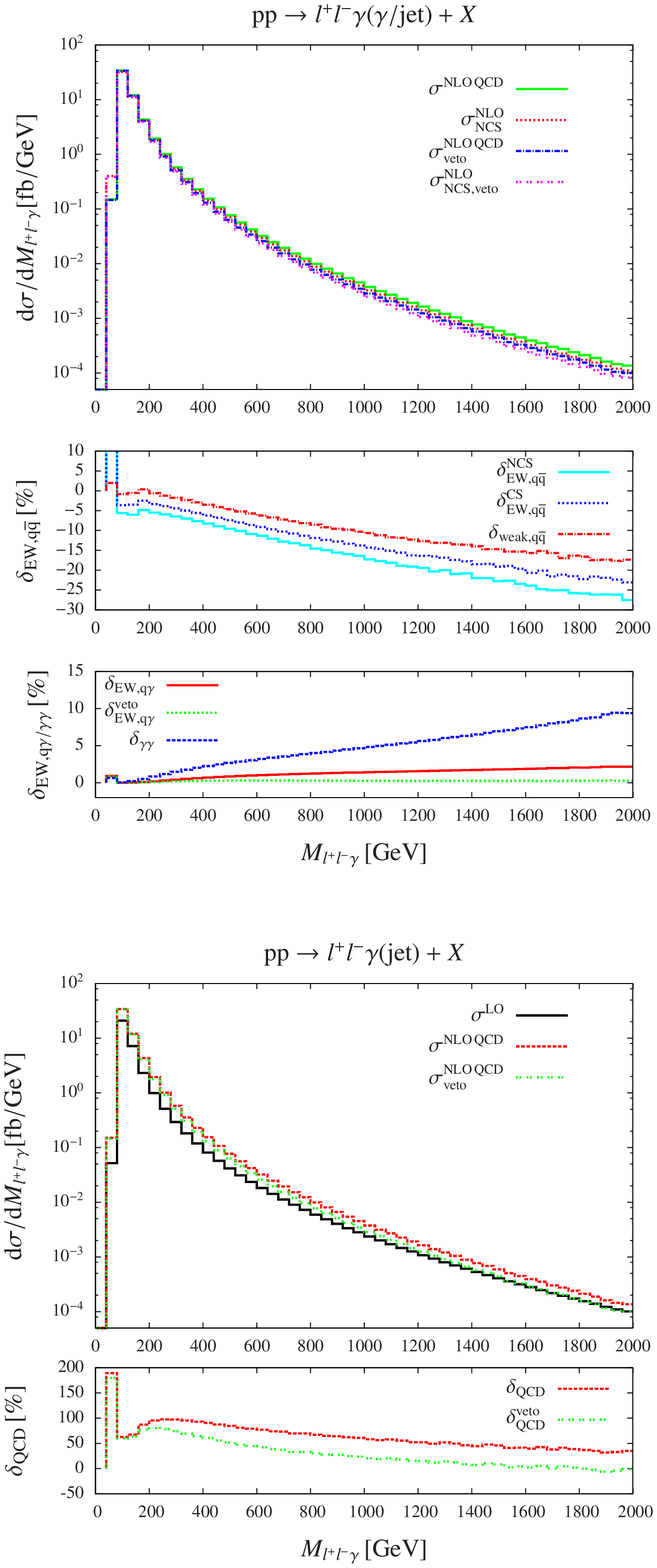}
 }
\caption{\label{fi:minv_zll} Distribution in the invariant mass
  $M_{\Plp\Plm}$ of the charged leptons (left) and distribution in the
  invariant three-body mass $M_{\Plp\Plm\gamma}$ of the charged
  leptons and the hardest photon (right), including EW (top) and QCD
  corrections (bottom). The large boxes show absolute predictions, the
  small ones relative corrections.}
\end{figure}

The invariant mass of the Z~boson, $M_{\Plp\Plm}$, is defined in \eqref{eq:Mllinv}, 
and the invariant three-body mass of the Z-decay products 
and the photon is defined by
\begin{align}
 M_{\Plp\Plm\gamma}& = \sqrt{\left(p_{\Plp}+p_{\Plm}+p_{\gamma_1}\right)^{2}}\, ,
\end{align}
where $p_{\Plp}$, $p_{\Plm}$, and $p_{\gamma_1}$ are the four-vectors of the 
charged leptons and the hardest photon, respectively. 
The corresponding distributions are shown in \reffi{fi:minv_zll}.
The invariant-mass distribution of the two charged leptons exhibits
two peaks already at LO. The larger one corresponds to the Z resonance
originating from the propagator that is resonant in the invariant mass
of the two charged leptons $M_{\Plp\Plm}$ at $M_{\Plp\Plm}=\MZ$.  The
smaller one comes from the resonance in the invariant three-body mass
$M_{\Plp\Plm\gamma}$, where the photon is radiated by one of the FS
charged leptons leading to a shift of the peak. The location of the
smaller peak mainly depends on the cut on the transverse momentum of
the photon.  With decreasing values of the cut on $p_{\mr{T},\gamma}$
the peak becomes less pronounced and moves towards the larger peak
until they fuse.  The QCD corrections are the leading corrections in
this distribution. They are particularly large at low invariant masses
and below the resonance with and without a jet veto.  This is to some
extent a result of our basic cuts, which allow invariant masses
$M_{\Plp\Plm}$ down to $40\GeV$, but at the same time demand
transverse momenta $p_{\mathrm{T},l^{\pm}}>25\GeV$. At LO, this leads
to a strong suppression of the cross section at low $M_{\Plp\Plm}$,
but at NLO QCD a jet recoil (with intermediate
$p_{\mr{T,jet}}<100\GeV$) in the real QCD corrections can lift such
events over the cuts on $p_{\mathrm{T},l^{\pm}}$, leading to
particularly large positive QCD corrections there.  In the resonance
region the EW corrections coming from the $\Pq\overline{\Pq}$~channel
are strongly dominated by photonic effects and reach $20\%$ in the CS
and $40\%$ in the NCS cases. Without photon recombination the shape
distortion of the Z resonance is larger, since more events appear
where the photon carries away energy and shifts events from higher to
lower energies.  The purely weak corrections are negligible in the
entire range we are looking at.  The quark--photon-induced EW
corrections are almost zero for low invariant masses and reach $1\%$
at $300\GeV$. In case of a jet veto they are well below one percent
everywhere.  The photon--photon-induced corrections are also tiny for
invariant masses below $100\GeV$, but grow up to $5\%$ at $300\GeV$.

Focusing on the invariant three-body mass we see that the QCD corrections are the dominating 
contribution in the region of low invariant masses, but decrease with and without a jet veto
to $50\%$ and $0\%$, respectively, for $M_{\Plp\Plm\gamma}=2\TeV$. 
In this region, the $M_{\Plp\Plm\gamma}$ distribution receives large
negative corrections up to $-18\%$ from the purely weak contribution,
and between $-23\%$ and $-28\%$ from the full EW corrections for the
CS and the NCS case, respectively.  The quark--photon-induced EW
corrections are of the order of $1$--$2\%$ and practically vanish 
in case of a jet veto, while the photon--photon-induced corrections
reach $10\%$ at $2\TeV$.  At high invariant mass the EW corrections
are of the same order of magnitude as the QCD corrections and become
the leading corrections in case of a jet veto.

\subsubsection{Results on rapidity and angular distributions}

In the following we present some rapidity and angular distributions
along with the corresponding NLO corrections. As for the integrated
cross section the QCD corrections typically yield the largest
contributions and in most cases a jet veto has no sizeable impact.  We
only show the most interesting distributions and do not single out the
purely weak corrections whenever they are negligibly small.
 
\begin{figure}
 \centerline{
         \includegraphics[scale=0.6]{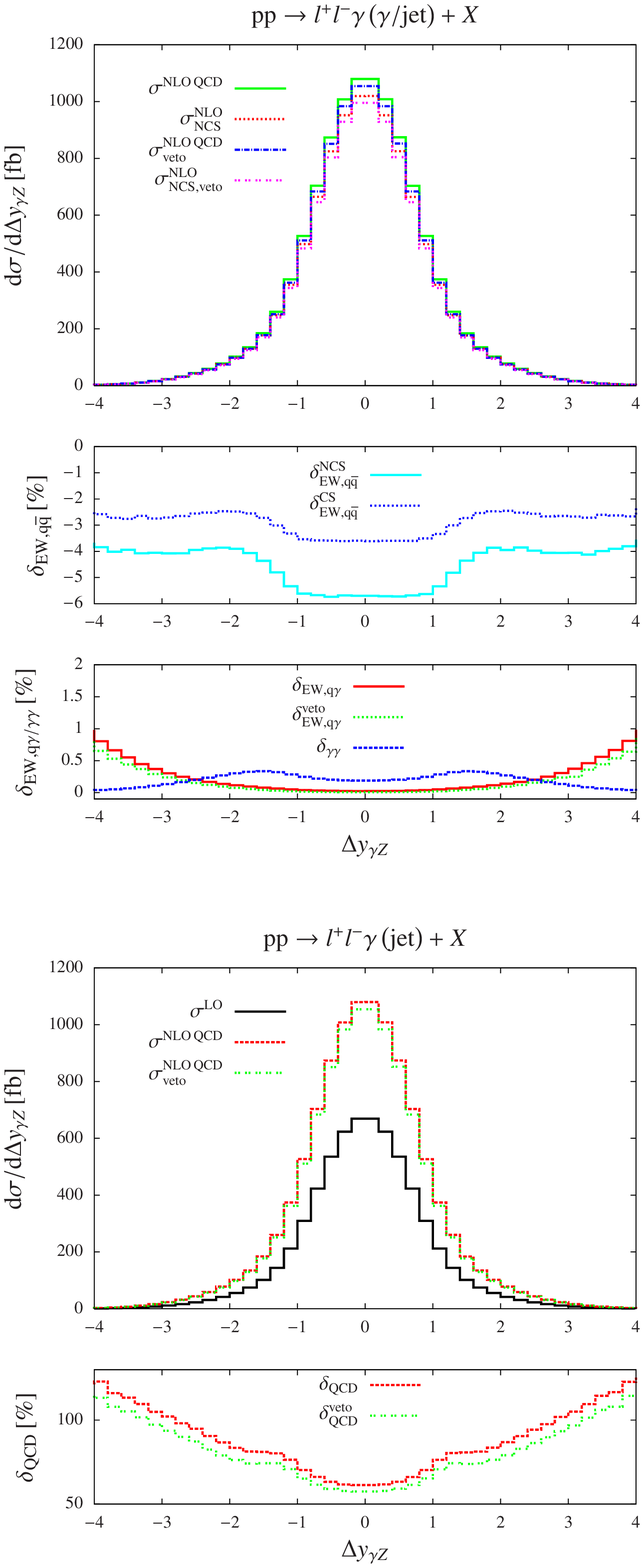}
 \hspace{-4em}
         \includegraphics[scale=0.6]{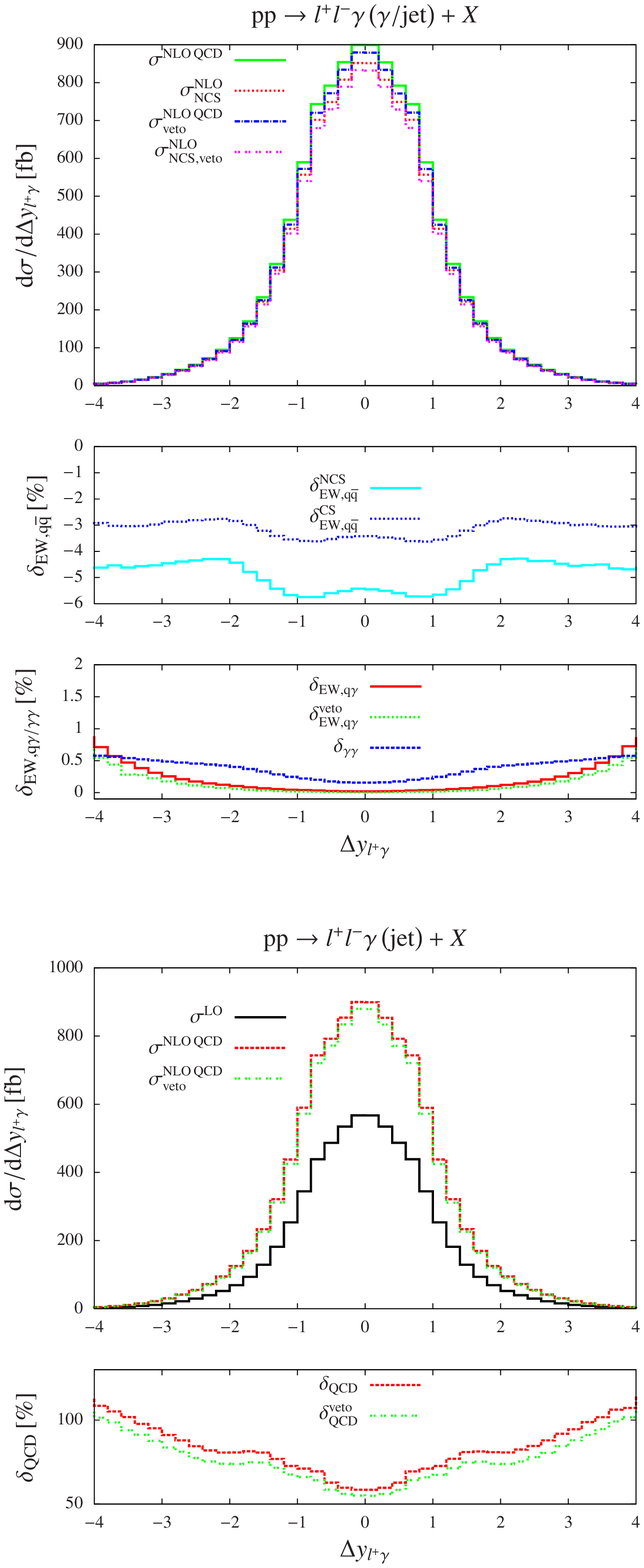}
 }
\caption{\label{fi:rap_gamma_Zll}Distributions in the
  rapidity difference $\Delta y_{\gamma \PZ}$ between the  hardest photon 
  and the Z~boson (left) and the rapidity difference $\Delta y_{\Plp\gamma}$ 
  between the charged lepton and the hardest photon (right), 
  including EW (top) and QCD corrections (bottom). The large boxes show
  absolute predictions, the small ones relative corrections.}
\end{figure}

Although the distributions in the rapidity differences $\Delta
y_{\gamma \PZ}$ and $\Delta y_{\Plp\gamma}$ shown in
\reffi{fi:rap_gamma_Zll} are different in their absolute values,
the relative QCD and EW corrections are very similar in the two cases.
The  QCD corrections
are about  $50\%$ at zero rapidity distance and grow to $110\%$ at
$|\Delta y| = 4$.  
The quark--antiquark-induced EW corrections amount to
roughly $-3\%$ in the CS case and vary between $-4\%$ and $-6\%$ in
the NCS case. The photon-induced corrections stay below
$1\%$ and are phenomenologically unimportant.
 
\begin{figure}     
 \centerline{
         \includegraphics[scale=0.6]{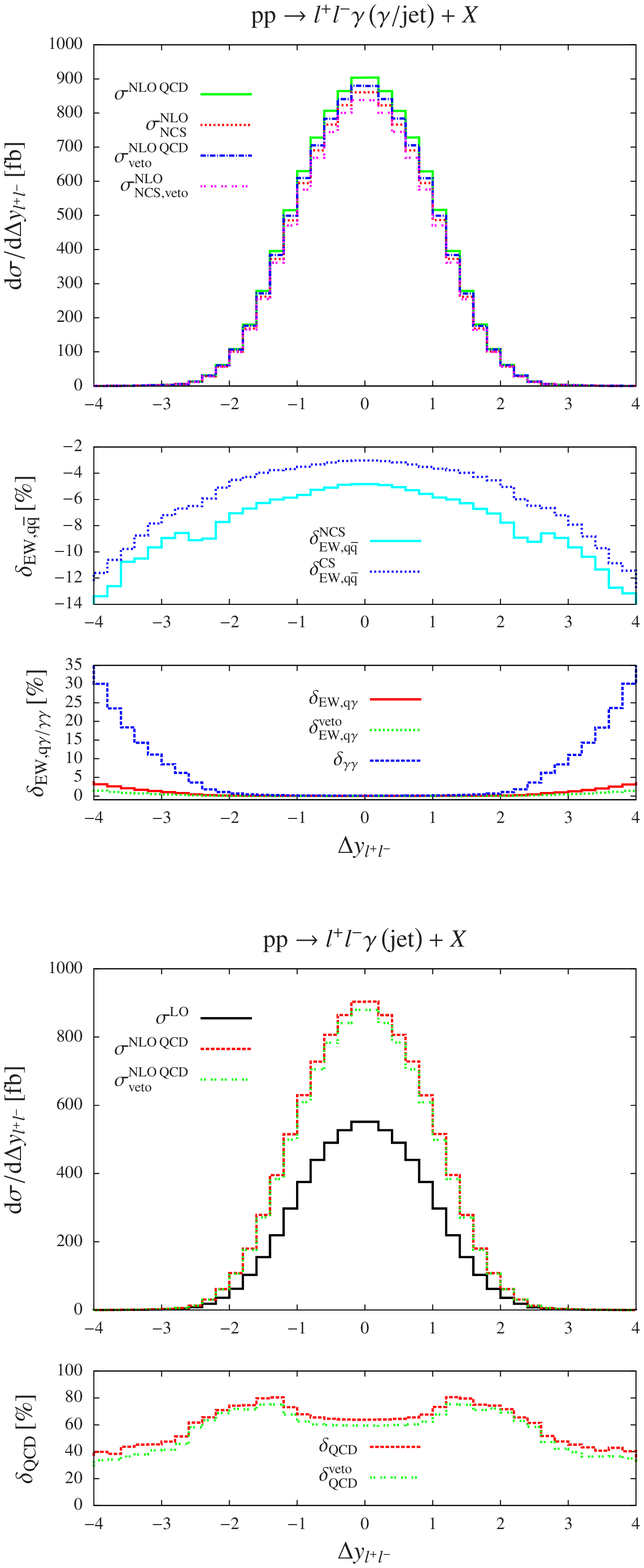}
 \hspace{-4em}
         \includegraphics[scale=0.6]{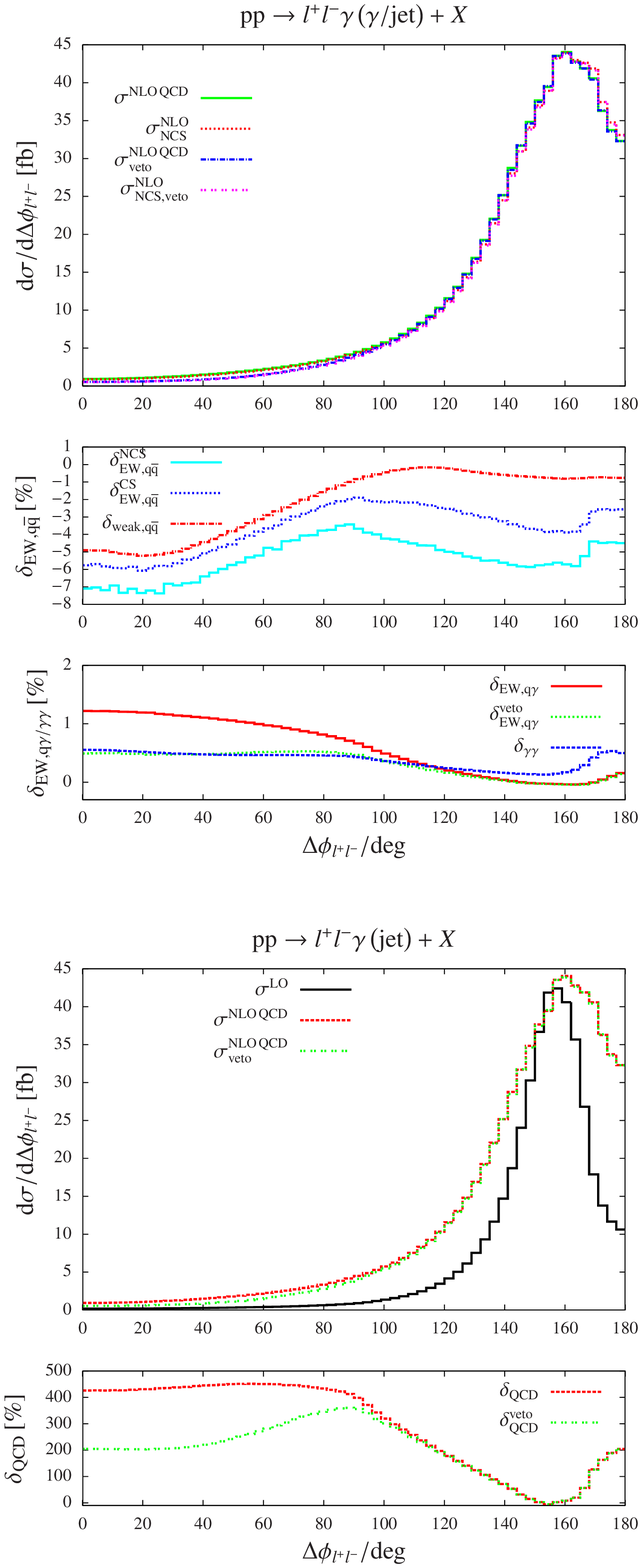}
 }
\caption{\label{fi:rap_angle_Zll} Distributions in the rapidity difference 
$\Delta y_{\Pl^+ \Pl^-}$ (left) and the azimuthal-angle difference 
$\Delta\phi_{\Pl^+ \Pl^-}$ (right) of the charged leptons,  
including EW (top) and QCD corrections (bottom). The large boxes show
  absolute predictions, the small ones relative corrections.}
\end{figure}

Next we focus on the rapidity difference and the azimuthal-angular
difference between the charged leptons shown in
\reffi{fi:rap_angle_Zll}. Starting with the rapidity difference we
see that the 
EW corrections to the $\Pq\overline{\Pq}$~channel have a minimum at zero
rapidity difference and increase up to $-12\%$ and $-14\%$ at $|\Delta
y_{\Plp\Plm}|=4$ in the CS and the NCS case, respectively.  The
corrections from the $\Pq\gamma$~channels are
below $4\%$ and $2\%$ with and without a jet veto, respectively.  The
photon--photon-induced corrections are below $5\%$ for $|\Delta
y_{\Plp\Plm}|< 2$ and increase steeply to $30\%$ for $|\Delta
y_{\Plp\Plm}|\sim 4$. However, in this region the cross section is
very small.

The azimuthal-angular difference between the charged leptons has a
peak around $160^{\circ}$.  This peak is caused by the cut on the
transverse momentum of the photon which eliminates events with
back-to-back leptons in the transverse plane. Increasing this cut
shifts the peak to smaller azimuthal angles.  The NLO QCD corrections
cause a very significant broadening of the peak, because jet recoil
effects strongly influence the angle between the leptons when the
decaying Z~boson receives a boost. The effect is strongest in the
limit where the leptons are nearly collinear, a region that is rarely
populated at LO, but receives large contributions from hard jet
emission where the jet recoil and the boost of the Z~boson are
strongest. This also explains the sensitivity of this region to the
jet veto.  The EW corrections from the $\Pq\Pqbar$~channels are of the
order of $-6\%$ and $-7\%$ in the CS and the NCS cases, respectively,
in the region of small angle differences and decrease at larger ones.
In this distribution the weak corrections are of the order of $-5\%$
at low angles and decrease to the $1\%$~level for angles around the
peak.  The photon-induced corrections lie below about $1\%$ and are
phenomenologically unimportant.

In summary, in angular and rapidity distributions the EW corrections are suppressed 
with respect to the QCD corrections.

\subsection{\boldmath{Invisible $\PZ+\gamma$ production: $\Pp\Pp \to
    \bar{\nu} \nu + \gamma + X$}} 
\label{se:zgamma_nn}

\subsubsection{Results on total cross sections}
\label{se:CSresults_zgamma_nn}

In \refta{ta:totcs_Znn} we present the LO cross sections
$\si^{\mr{LO}}$ for different $\Pp\Pp$ centre-of-mass energies
$\sqrt{s}$ and different types of relative corrections $\de$ defined
in \refeq{eq:relcor} for $\Pp\Pp\to \bar{\nu} \nu + \gamma + X$.
\begin{table}
  \centering\renewcommand{\arraystretch}{1.2}
  \begin{tabular}[H]{c|rrr}
    \multicolumn{4}{c}{$\Pp\Pp \to \bar{\nu} \nu  \gamma + X $} \\
    \hline\hline
    $\sqrt{s}/\TeV\; $ & $7$\phz & $8$\phz & $14$\phz \\
    \hline\hline
    $\si^{\mr{LO}}             /\fba\;$ &$\; 74.927(2)$&$\; 91.031(1)$&$\; 185.254(4)$  \\ 
    \hline \hline
    $\de_{\phot, \Pq\overline{\Pq}}/\% \;$   &$\;  0.30$\phz &$\;  0.30$\phz &  $\;  0.29(1)$     \\
    \hline
    $\de_{\mr{weak}, \Pq\overline{\Pq}}/\% \;$   &$\;  -4.45$\phz &$\;  -4.56$\phz &  $\;  -4.98$\phz     \\
    \hline \hline
    $\de^{\mr{frag}}_{\EW, \Pq\ga}/\% \;$        &$\;  0.03$\phz &$\; 0.04$\phz &$\;  0.03$\phz         \\ 
    \hline 
    $\de^{\mr{veto},\,\mr{frag}}_{\EW, \Pq\ga}/\% \;$ &$\; 0.02$\phz &$\;  0.03$\phz &$\;  0.02$\phz         \\ 
    \hline
    $\de^{\mr{Frix}}_{\EW, \Pq\ga}/\% \;$        &$\; 0.03$\phz &$\; 0.03$\phz &$\; 0.02$\phz         \\ 
    \hline 
    $\de^{\mr{veto},\,\mr{Frix}}_{\EW, \Pq\ga}/\% \;$ &$\;  0.02$\phz &$\; 0.02 $\phz &$\;  0.01$\phz         \\ 
    \hline \hline
    $\de^{\mr{frag}}_{\QCD}/\% \;$        &$\;  46.35(4)$ &$\;  46.94(5)$ &$\;  51.59(5)$    \\ 
    \hline
    $\de^{\mr{Frix}}_{\QCD}/\% \;$        &$\;45.46(4)  $ &$\;  46.07(5)$ &$\;  50.66(3)$    \\ 
    \hline
    $\de^{\veto,\,\mr{frag}}_{\QCD}/\% \;$           &$\;  42.57(4)$ &$\;  42.54(3)$ &$\;  44.11(3)$     \\ 
    \hline
    $\de^{\veto,\,\mr{Frix}}_{\QCD}/\% \;$     &$\;41.71(4)  $ &$\; 41.67(3)$ &$\;  43.28(3)$     \\ 
    \hline \hline
  \end{tabular}
  \caption{\label{ta:totcs_Znn} 
    Integrated cross sections  and relative corrections for $\Pp\Pp
    \to \bar{\nu} \nu\gamma + X $  
    at different LHC energies. The EW corrections to the quark--antiquark 
    annihilation channels are split into purely weak and photonic corrections.
    Contributions from the
    photon-induced channels and QCD corrections are shown with a jet
    veto (veto) as well as without a jet veto
    using a fragmentation function (frag) or
    the Frixione isolation criterion (Frix) to separate photons and jets.
    The numbers in
    parentheses denote the integration errors in the last digits.  This
    error is omitted if it is negligible at the given accuracy.} 
\end{table}
Recall that we sum over all three lepton generations. Similar to the
results in \refta{ta:totcs_Zll} we find that the relative corrections
only marginally vary for the different collider energies. Here again
the QCD corrections give the dominant contributions with
$\sim40{-}50\%$, about a third to a half of which results from the
gluon-induced channels (not separately shown), a much larger share
than for $l^+l^-\ga$ production. Owing to the neutral final state the
dominant contribution inside the quark--antiquark-induced EW
corrections results from pure weak corrections with $\sim -5\%$ and
the photonic corrections only contribute $0.3\%$. Again the
quark--photon-induced corrections are phenomenologically negligible.

\subsubsection{Results on transverse-momentum distributions}
\label{se:results_mom_distr_zgamma_nn}

\begin{figure}     
 \centerline{
         \includegraphics[scale=0.6]{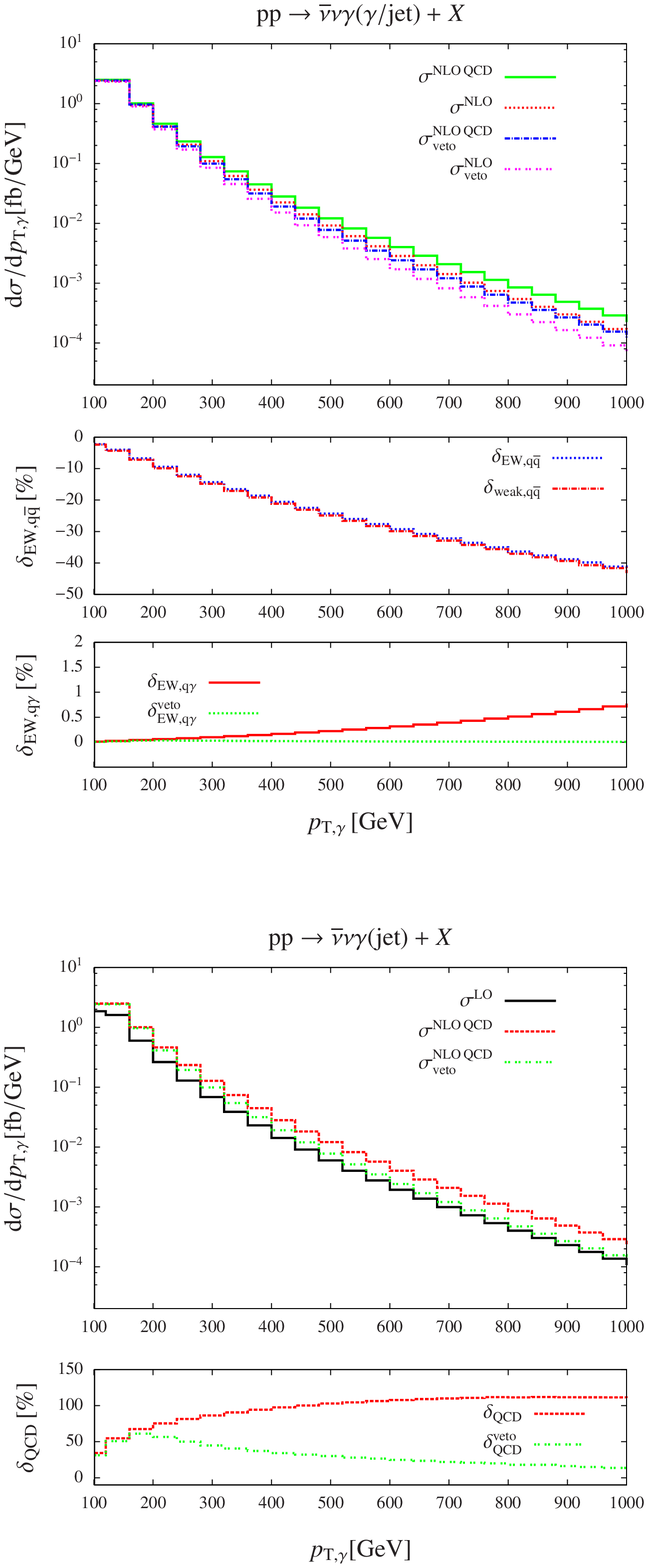}
 \hspace{-4em}
         \includegraphics[scale=0.6]{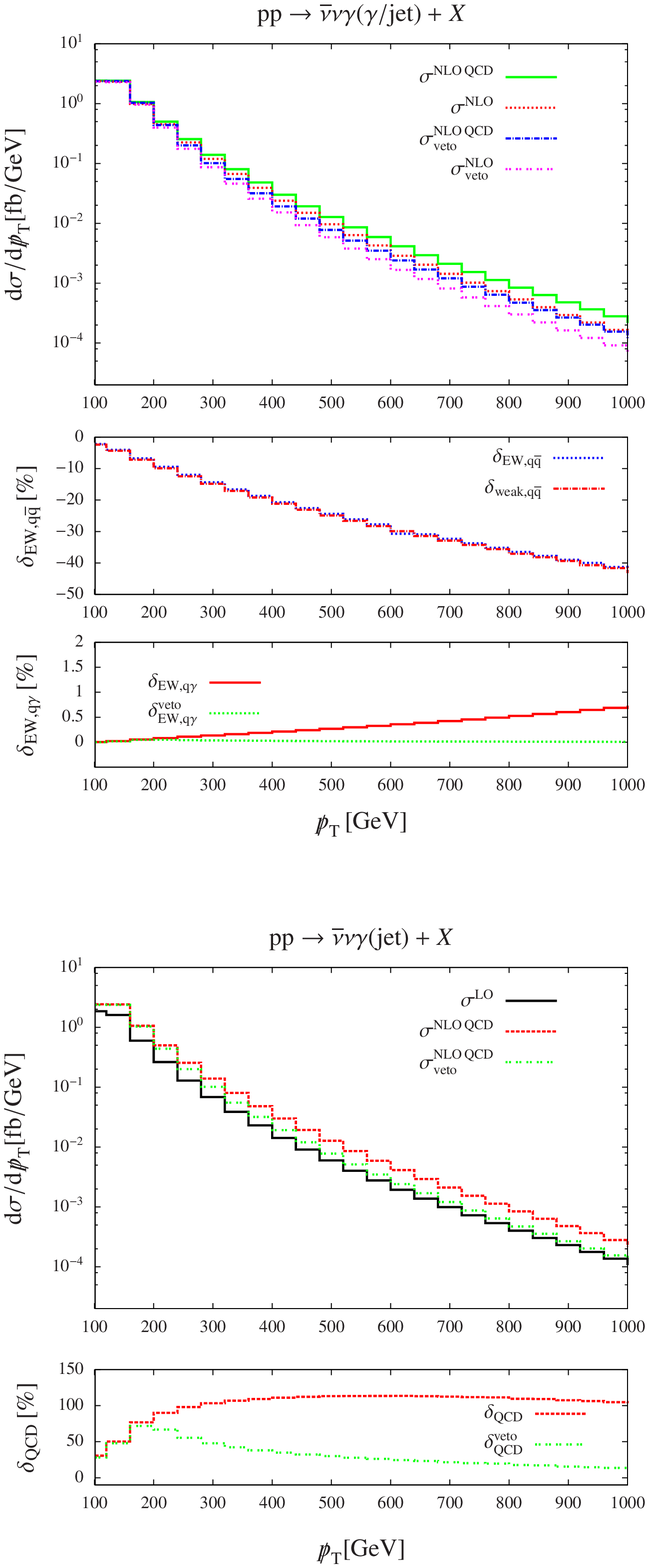}
 }
\caption{\label{fi:pt_boson_znn} Distributions in the
  transverse momentum $p_{\mathrm{T}}$ of the photon (left) and the
  missing transverse momentum (right), including EW (top) and QCD
  corrections (bottom). The large boxes show absolute predictions, the
  small ones relative corrections.}
\end{figure}

In the following we present differential distributions including QCD
and EW corrections to $\Pp\Pp\to \bar\nu \nu \gamma + X$ for a
$\Pp\Pp$ centre-of-mass energy of $14 \TeV$.  In
\reffi{fi:pt_boson_znn} we show distributions in the transverse
momentum of the photon and in the missing transverse momentum. First
we notice that the two distributions as well as the corresponding
corrections are almost identical.
Since the photon neither couples to the Z~boson nor to the neutrinos,
the photon and the Z~boson are always back to back in their
centre-of-mass frame at LO.  Corrections from the real radiation of
jets or photons off the initial-state partons hardly distinguish
between the produced Z~boson or the hard photon, so that even the NLO
corrections (both QCD and EW) almost coincide for the $p_{\mr{T},\PZ}$ and
$p_{\mr{T},\gamma}$ distributions.
Furthermore the NLO corrections closely resemble
the ones shown in \reffi{fi:pt_boson_zll} (left) for the
$p_{\mr{T},\gamma}$ distribution for the $l^+l^-\gamma$ final state. The QCD corrections are similar, because 
they only affect the IS quarks and do not depend on the final state. The  
EW corrections corresponding to the $\Pq\overline{\Pq}$~channel are 
identical with the weak corrections including the large 
Sudakov logarithms and turn out to be of similar size quite
independent of the final state. 
The photonic corrections, which only involve the IS quarks, are negligible for
$\bar\nu\nu\gamma$ production,
i.e.~they are almost completely absorbed into the PDFs.  
The quark--photon-induced corrections roughly differ by a factor of two
in the cases of $l^+l^-\gamma$ and $\bar\nu\nu\gamma$ production, 
since they depend on the FS particles: In the visible decay channel the IS photon (discussed in 
\refse{se:zgamma_ll}) can also couple to the FS charged leptons,
whereas in the invisible decay channel it can only couple to the IS quarks.

\subsubsection{Results on transverse-mass distributions}
\label{se:results_mass_distr_zgamma_nn}

\begin{figure}
 \centerline{
         \includegraphics[scale=0.6]{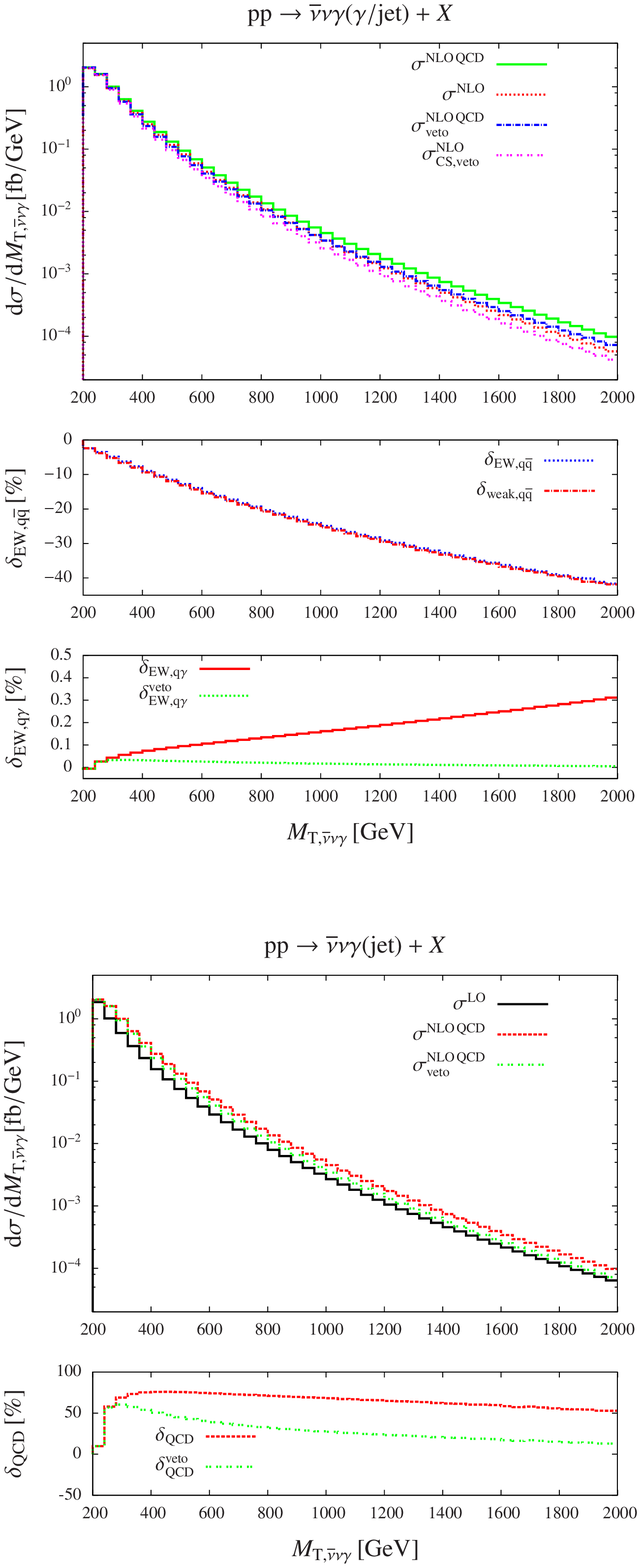}
 \hspace{-4em}
         \includegraphics[scale=0.6]{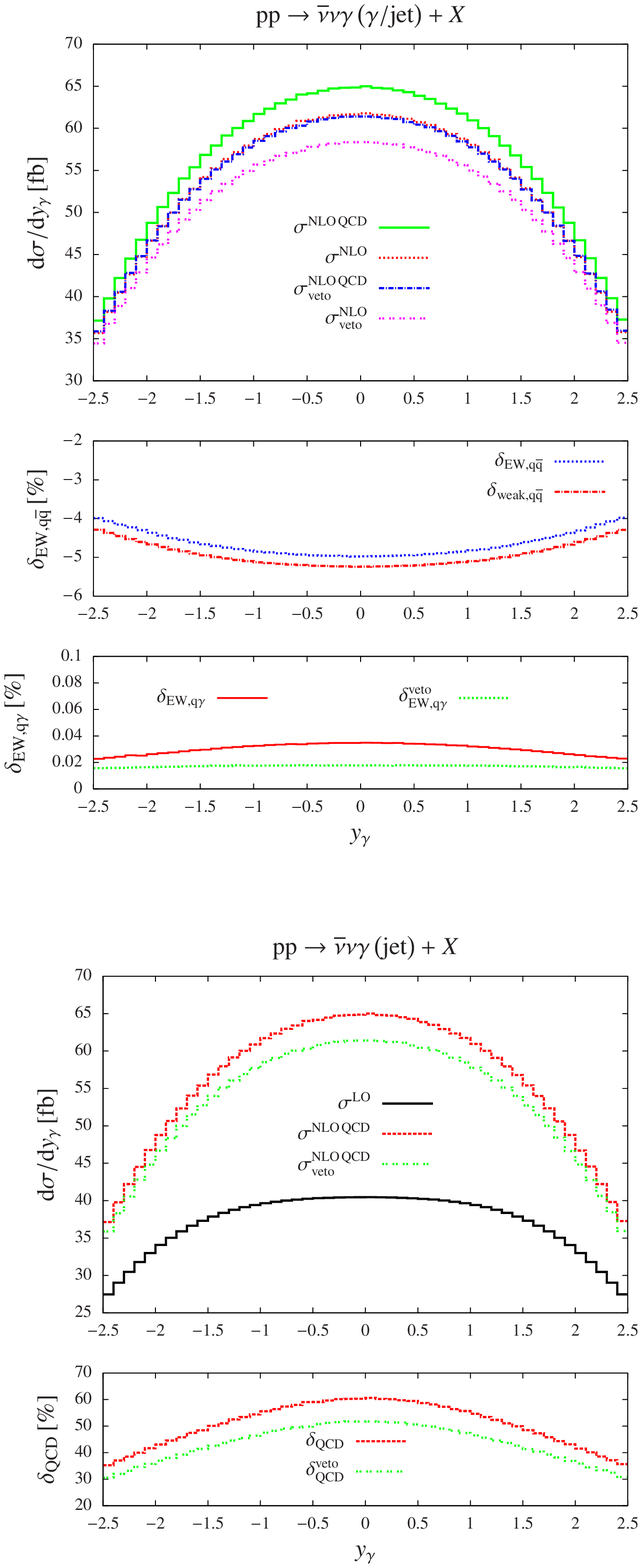}
 }
\caption{\label{fi:mt_znn} Distributions in the transverse three-body mass 
  $M_{\mr{T},\bar{\nu}\nu\gamma}$ of the neutrino pair and the hardest
  photon and in the rapidity $y_{\gamma}$ of the hardest photon,
  including EW (top) and QCD corrections (bottom). The large boxes
  show absolute predictions, the small ones relative corrections.}
\end{figure}

The transverse three-body mass of the neutrinos and the photon is given by
\begin{align}
   M_{\mr{T},\,\bar{\nu}\nu\gamma}& 
=  \sqrt{(\slashed{p}_{\mathrm{T}}+p_{\mr{T}, \gamma_1})^{2}
- (\slashed{\bf p}_{\mathrm{T}}+{\bf p}_{\mr{T}, \gamma_1})^{2}
}
\, ,
\end{align}
where we always take the hardest photon if there are two.
The corresponding distribution is shown on the left side of \reffi{fi:mt_znn}. 
Comparing 
this with the invariant three-body mass of the charged leptons and the photon
given in \reffi{fi:minv_zll}, we see that the QCD corrections are flat and 
have the same trend in both distributions. This can be explained with the same 
argument as in case of the transverse-momentum distributions, since the 
QCD corrections only act on the IS quarks and do not depend on the FS leptons. 
Note that in the invisible decay channel the distribution 
only starts at $190\GeV$ at NLO and at   $200\GeV$ at LO
owing to the larger $p_{\mr{T}}$~cuts. The EW corrections to the
$\Pq\Pqbar$~channel are considerably larger in the invisible channel
which is due to the fact that we consider the {\it transverse}
three-body mass instead of the full three-body mass. If the latter
gets large, there is still the possibility that all transverse momenta
are moderate or small. By contrast a large transverse three-body mass
requires some large transverse momenta, so that the kinematical
configuration is closer to the Sudakov regime where all Minkowski
products of momenta are large and EW corrections are strongly
enhanced.  The corrections from the quark--photon channel are below
$1\%$ with and without jet veto and therefore negligible.

\subsubsection{Results on rapidity distributions}
\label{se:results_rapidity_distr_zgamma_nn}

The rapidity distribution of the hardest photon is shown on the right
side of \reffi{fi:mt_znn}.  It receives large QCD corrections between
$30\%$ and $60\%$. The jet veto diminishes the QCD corrections by
$5{-}10\%$. The EW corrections to the $\Pq\Pqbar$~channel mainly
originating from the purely weak corrections are of the order of
$-5\%$ and almost flat and therefore reflecting the corrections to the
integrated cross section.  The EW corrections are small compared to
the QCD corrections, but not completely negligible.

\subsection{Results with anomalous triple gauge-boson couplings}
\label{se:AC}

In order to parametrize effects of new physics influencing the non-abelian 
gauge-boson couplings, higher-dimensional operators can be added to
the SM Lagrangian.
The commonly used form of anomalous triple gauge-boson couplings
(aTGCs) goes back to \citere{Hagiwara:1986vm} and is based on a
general parametrization of the $\PW\PW V$, $\PZ\PZ V$, and $\PZ\gamma
V$ vertices (assuming that $\PW$ and $\PZ$~bosons couple to conserved
currents), with $V=\PZ,\gamma$.  In the following we employ the
definition of the aTGCs following
\citeres{Gounaris:1996rz,Gounaris:1999kf}.

The case of anomalous $\PZ\gamma V$ ($V=\PZ,\gamma$) couplings is
particularly interesting, since they do not appear as elementary
couplings in the SM.  Following \citere{Gounaris:1999kf}, we assume
Lorentz and $\mr{U}(1)_{\mr{em}}$ gauge invariance as well as Bose
symmetry. With these assumptions the most general Lagrangian that
describes the anomalous $VVV$ vertex is given by
\begin{eqnarray}
\label{eq:anLag-ZAV}
{\cal{L}}_{VVV} &=& \frac{e}{\MZ^{2}} \Bigg [
-[f_4^\gamma (\partial_\mu F^{\mu \beta})-
f_4^Z (\partial_\mu Z^{\mu \beta}) ] Z_\alpha
( \partial^\alpha Z_\beta)\nonumber\\
& & \qquad\;{}+[f_5^\gamma (\partial^\sigma F_{\sigma \mu})-
f_5^Z (\partial^\sigma Z_{\sigma \mu}) ] \widetilde{Z}^{\mu \beta} Z_\beta
\nonumber \\
&&\qquad\;{}+  [h_1^\gamma (\partial^\sigma F_{\sigma \mu})
-h_1^Z (\partial^\sigma Z_{\sigma \mu})] Z_\beta F^{\mu \beta}
+[h_3^\gamma  (\partial_\sigma F^{\sigma \rho})
- h_3^Z  (\partial_\sigma Z^{\sigma \rho})] Z^\alpha
 \widetilde{F}_{\rho \alpha}
\nonumber \\
&&\qquad\;{}+ \left \{\frac{h_2^\gamma}{\MZ^{2}} [\partial_\alpha \partial_\beta
\partial^\rho F_{\rho \mu} ]
-\frac{h_2^Z}{\MZ^{2}} [\partial_\alpha \partial_\beta
(\square +\MZ^{2}) Z_\mu] \right \} Z^\alpha F^{\mu \beta}
\nonumber \\
&&\qquad\;{}- \left \{
\frac{h_4^\gamma}{2\MZ^{2}}[\square \partial^\sigma
F^{\rho \alpha}] -
\frac{h_4^Z}{2 \MZ^{2}} [(\square +\MZ^{2}) \partial^\sigma
Z^{\rho \alpha}] \right \} Z_\sigma \widetilde{F}_{\rho \alpha }
 \Bigg ] ~ ,
\end{eqnarray}
where $Z^{\mu}$ is the $\PZ$-boson field, $Z^{\mu\nu}=\partial^\mu
Z^\nu -\partial^\nu Z^\mu$, $\widetilde{Z}_{\mu \nu}=\epsilon_{\mu \nu
  \rho \sigma}Z^{\rho\sigma}/2$, 
and $\widetilde{F}_{\mu \nu}=\epsilon_{\mu \nu \rho
  \sigma}F^{\rho\sigma}/2$.  The anomalous couplings proportional to
$f_4^V, ~h_1^V,~ h_2^V$ violate CP symmetry whereas the ones
proportional to $f_5^V$, $h_3^V$, $h_4^V$ respect it.  Note that our
conventions for the SM Lagrangian taken from \citere{Denner:1991kt}
differ from those of \citere{Gounaris:1999kf} by a minus sign in the
Z-boson and other fields not appearing in Eq.~\eqref{eq:anLag-ZAV}, 
a difference that uniformly applies to SM and
non-standard couplings.  The operators in Eq.~\eqref{eq:anLag-ZAV}
exploit all possible Lorentz structures that do not include the scalar
components of any of the two vector bosons, i.e.\ the Lagrangian
assumes that
\begin{equation}
\partial_\mu A^\mu=0, \qquad  \partial_\mu Z^\mu=0.
\end{equation}
This relation also effectively holds for virtual photons and Z~bosons
in our case, since terms containing $\partial \PZ$ lead to
contributions to amplitudes that are proportional to the lepton or
quark masses, which are neglected, because the Z~boson couples to a
conserved current in the limit of small fermion masses.  With the
Lagrangian \eqref{eq:anLag-ZAV} the momentum-space Feynman rules for
the anomalous $\PZ\PZ\gamma$ and $\PZ\gamma\gamma$ vertices read
\begin{flalign}
\label{eq:atgc-zza}
\Gamma^{\mu\nu\alpha}_{\PZ\PZ\gamma}(Q,q,k)=&-\frac{\mr{i}e}{\MZ^{2}}\klam{Q^{2}-q^{2}}\left\{h^{\PZ}_{1}\klam{k^{\mu}g^{\nu\alpha}-k^{\nu}g^{\mu\alpha}} 
                                                                            -h^{\PZ}_{3}\epsilon^{\mu\nu\alpha\beta} k_{\beta}\right\}\nonumber\\
                                    &-\frac{\mr{i}e}{\MZ^{2}}\klam{Q^{2}-\MZ^{2}}
                                     \left\{\frac{h^{\PZ}_{2}}{\MZ^{2}}Q^{\nu}\left[\klam{k\cdot Q}g^{\mu\alpha}-k^{\mu}Q^{\alpha}\right]
-\frac{h^{\PZ}_{4}}{\MZ^{2}}Q^{\nu}\epsilon^{\mu\alpha\beta\delta} Q_{\beta}k_{\delta}\right\}\nonumber\\
                                    &-\frac{\mr{i}e}{\MZ^{2}}\klam{q^{2}-\MZ^{2}}
                                     \left\{\frac{h^{\PZ}_{2}}{\MZ^{2}}q^{\nu}\left[\klam{k\cdot q}g^{\nu\alpha}-k^{\nu}Q^{\alpha}\right]
-\frac{h^{\PZ}_{4}}{\MZ^{2}}q^{\mu}\epsilon^{\nu\alpha\beta\delta} q_{\beta}k_{\delta}\right\}+\ldots,\\
\Gamma^{\mu\nu\alpha}_{\PZ\gamma\gamma}(Q,q,k)=&\frac{\mr{i}e}{\MZ^{2}}q^{2}\bigg\{-h^{\gamma}_{1}\klam{k^{\mu}g^{\nu\alpha}-k^{\nu}g^{\mu\alpha}} 
                                                                            +h^{\gamma}_{3}\epsilon^{\mu\nu\alpha\beta} k_{\beta}\bigg.\nonumber\\
                                         &{}+\bigg.\frac{h^{\gamma}_{2}}{\MZ^{2}}q^{\mu}\left[\left(k\cdot q\right)g^{\nu\alpha}-k^{\nu}q^{\alpha}\right]
                                            -\frac{h^{\gamma}_{4}}{\MZ^{2}}q^{\mu}\epsilon^{\nu\alpha\delta\rho}q_{\delta}k_{\rho}\bigg\} + \ldots,
\end{flalign}
where all momenta are considered as incoming and all terms are omitted
that do not contribute for an on-shell photon with momentum $k$.  
Assuming that one
Z~boson is approximately on shell ($q^{2}\sim \MZ^{2}$), we find the
same vertex as derived in
\citeres{Gounaris:1999kf,DeFlorian:2000sg}.

The anomalous couplings spoil unitarity of the S-matrix in the
limit of high energies. This behaviour is usually tamed by including
form factors, mimicking the onset of new physics
that damps the effects of the aTGCs at high momentum transfer.
We use the standard form factors
\begin{align}
\label{eq:formfactor}
h^{V}_{i}\rightarrow \frac{h^{V}_{i}}{\left(1+\frac{M^{2}_{\PZ\gamma}}{\La^{2}}\right)^{n}}\, ,
\end{align}
where $V=\gamma,\PZ$, the scale of new physics is denoted as $\La$, and $M_{\PZ\gamma}$ is
the invariant mass of the Z-boson--photon system. The exponent $n$
is chosen such that the form factor decreases fast enough to restore
unitarity. 

In order to combine the contribution of the anomalous couplings (AC)
with the NLO corrections in a consistent way, we 
extend Eq.~\eqref{eq:naive-product}
by the relative anomalous contribution $\delta_{\mr{AC}}$,
\begin{align}\label{eq:naive-product-ac}
    \sigma^{\mr{NLO}}_{\mr{AC}}&= \sigma^{\mr{NLO\,QCD}}
                     \klam{1+\delta_{\EW,\Pq\overline{\Pq}}+\delta_{\mr{AC}}}
                     +\Delta\sigma^{\mr{NLO\,EW}}_{\Pq\gamma}+\klam{\Delta\sigma^{\mr{NLO\,EW}}_{\gamma\gamma}},
\end{align}
where $\delta_{\mr{AC}}$ is defined by
\begin{align}
  \delta_{\mr{AC}}&= \frac{\sigma^{\NLO\,\QCD}_{\mr{AC}}}{\sigma^{\NLO\,\QCD}}-1 \, .
\end{align}
The SM cross section $\sigma^{\NLO\,\QCD}$ is defined in Sect.~\ref{se:setup},
and $\si^{\NLO\,\QCD}_{\mr{AC}}$ is the NLO QCD cross section including the aTGC contribution.
Thus, $\delta_{\mr{AC}}$ can be considered as an additional
correction on top of the EW correction in \eqref{eq:naive-product}
which we choose to combine linearly. A proper combination of aTGCs 
and EW corrections would require an effective-field-theory 
approach, which goes beyond the scope of this work.
In contrast, QCD corrections can be calculated in a straightforward
way in the presence of aTGCs.

For our calculation we choose values for the ACs consistent with
the most recent limits set by the ATLAS and 
CMS collaborations in \citeres{Aad:2013izg,Khachatryan:2015kea}.
Following these references 
we demand CP conservation which is equivalent to $h^{V}_{1,2}=0$. 
In \citeres{Aad:2013izg,Khachatryan:2015kea} limits were set on the 
remaining ACs for two different scales that enter the form factor defined in 
Eq.~\eqref{eq:formfactor}. We choose the following two sets of values,
\begin{align}
\Lambda=3\TeV:\qquad  h^{\gamma}_{3} = 2.4\cdot 10^{-2},\qquad  h^{\gamma}_{4} = 3.6\cdot 10^{-4},
              \qquad  h^{\PZ}_{3}= 2.0\cdot 10^{-2}, \qquad h^{\PZ}_{4}= 3.1\cdot 10^{-4}; \nonumber\\
\Lambda\rightarrow\infty:\qquad  h^{\gamma}_{3} = 4.6\cdot 10^{-3},\qquad  h^{\gamma}_{4} = 3.5\cdot 10^{-5},
                    \qquad h^{\PZ}_{3}= 3.7\cdot 10^{-3}, \qquad h^{\PZ}_{4}= 3.0\cdot 10^{-5}.
\label{eq:AC}
\end{align}
The former numbers for $\Lambda=3\TeV$ reflect the limits set by ATLAS~\cite{Aad:2013izg}
using data from the run at an energy of $7\TeV$ with a luminosity of $4.6\fba^{-1}$,
the latter values without form factor ($\Lambda\rightarrow\infty$)
correspond to the limits set by CMS~\cite{Khachatryan:2015kea}
after collecting a luminosity of $19.5\fba^{-1}$ at $8\TeV$.
Following \citere{DeFlorian:2000sg} we choose the exponent of the form factor 
as $n=3$ and $n=4$ for the ACs $h^{V}_{3}$ and $h^{V}_{4}$, respectively. 

Analyzing the impact of aTGCs for $\sqrt{s}=14\TeV$ we only
present results obtained without a jet veto, since the impact of a jet
veto does not change the effect of the ACs significantly.  Note that
we only present QCD-corrected distributions in the following.
Therefore we do not have to distinguish between the CS and the NCS
cases.

\subsubsection{$\Pp\Pp\to \Pl^+ \Pl^- + \gamma + X$}
\begin{figure}     
 \centerline{
 \hspace{-3em}
         \includegraphics[scale=0.6]{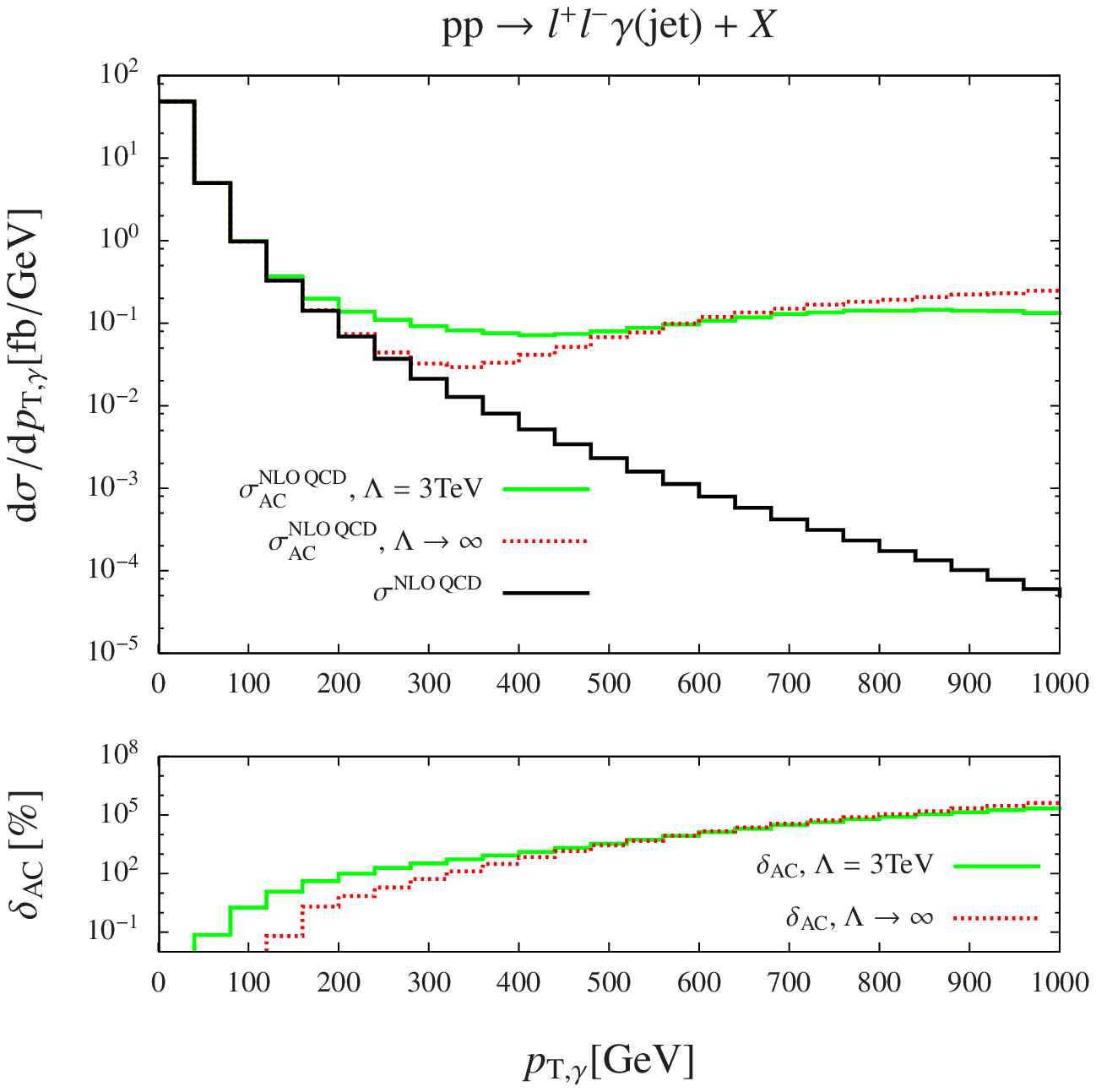}
 \hspace{-3em}
         \includegraphics[scale=0.6]{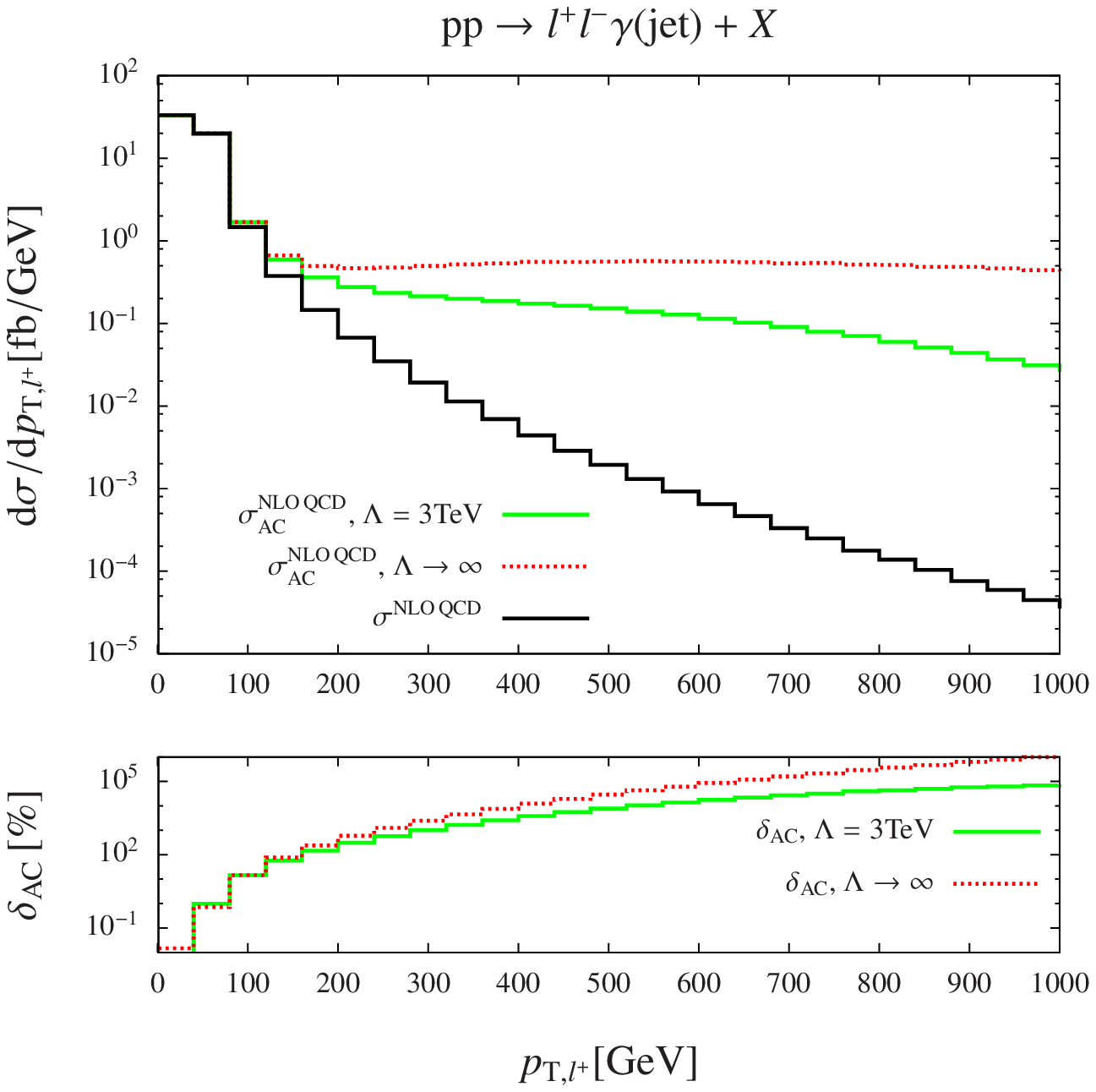}
 }
\caption{\label{fi:pt_zll_ac} Absolute and relative contributions of
  aTGCs to the transverse-momentum distributions of the photon (left)
  and the charged lepton (right).}
\end{figure} 
In \reffi{fi:pt_zll_ac} we analyse the impact of aTGCs on the
transverse-momentum distributions of the photon and the charged lepton
corresponding to the visible decay channel of the $\PZ$~boson.
Focusing on the $p_{\mr{T},\gamma}$ distribution we see that the aTGCs
start to cause a visible effect roughly at $200\GeV$ and at $350\GeV$
in case of $\Lambda=3\TeV$ and $\Lambda\rightarrow\infty$,
respectively. The relative contributions of the aTGCs meet at
$450\GeV$ and develop in the same way staying almost constant. The
relative corrections are huge growing up to $10^{3}$ at $1\TeV$.  The
situation is different in the $p_{\mr{T},\Plp}$ distribution. Here the
contributions of the aTGCs from the two setups overlap at small
transverse momenta and start to have a visible effect around
$150\GeV$. At higher transverse momenta they diverge, where the
contribution without form factor remains almost constant, whereas the
contribution with form factor decreases. The relative impact coming
from the aTGCs reach a factor of $10^{4}$ at $1\TeV$.  For a fixed set
of AC values, one of course would expect larger aTGC effects for the
case without form factor $\Lambda\to\infty$, since a finite form
factor effectively switches off the AC contribution at high energies.
Recall, however, that our AC values chosen for $\Lambda\to\infty$
correspond to limits set in a fit to data collected at a somewhat
higher pp energy with a significantly higher luminosity, so that at
least for the formerly experimentally accessible energy scales in the
distributions the impact of the ACs with $\Lambda\to\infty$ is
expected to be somewhat smaller than for the set of AC values with
$\Lambda=3\TeV$.  This behaviour is, for instance, found in the
$p_{\mr{T},\gamma}$ distribution in \reffi{fi:pt_zll_ac}.

\begin{figure}
 \centerline{
 \hspace{-3em}
         \includegraphics[scale=0.6]{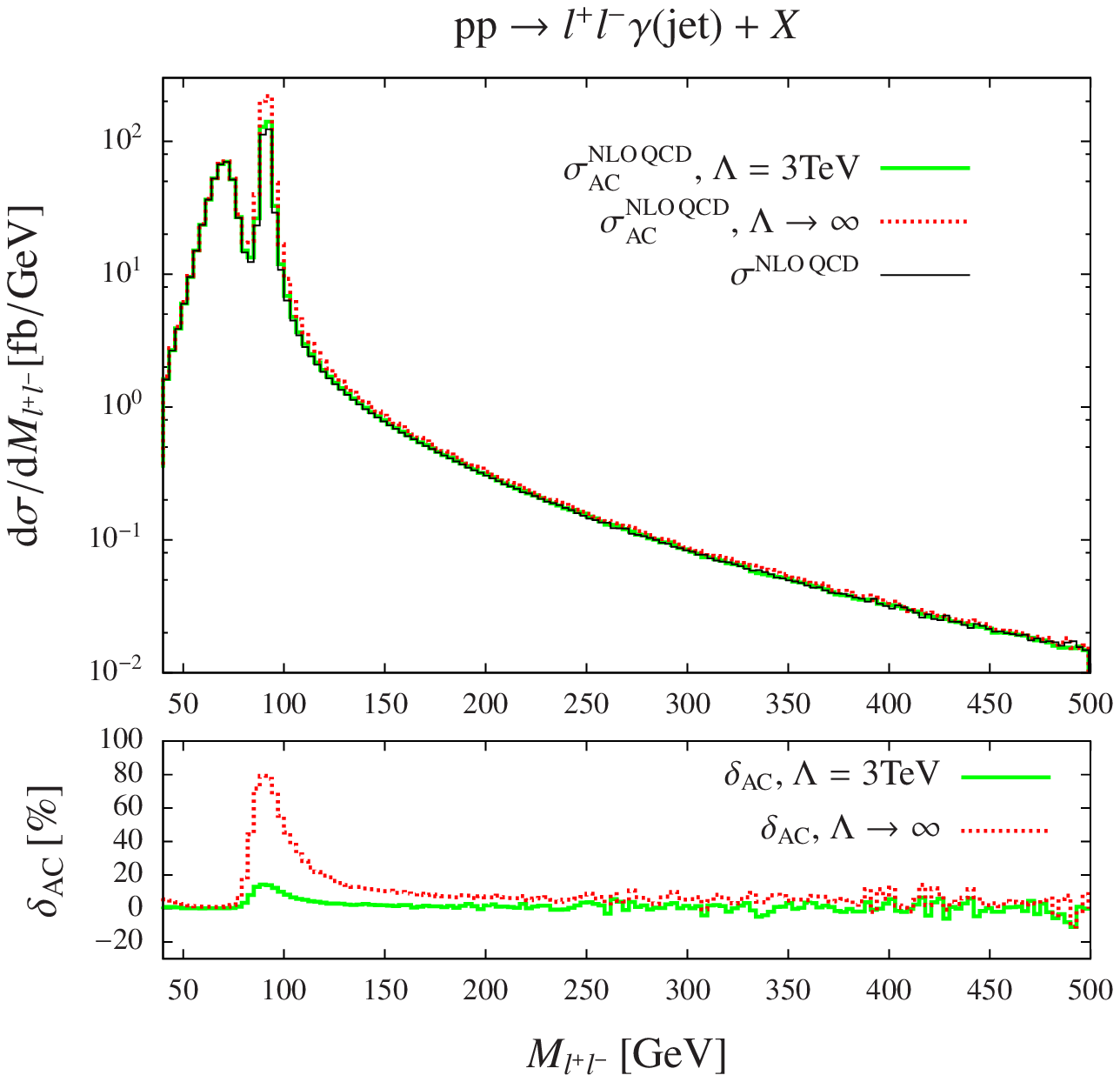}
 \hspace{-3em}
         \includegraphics[scale=0.6]{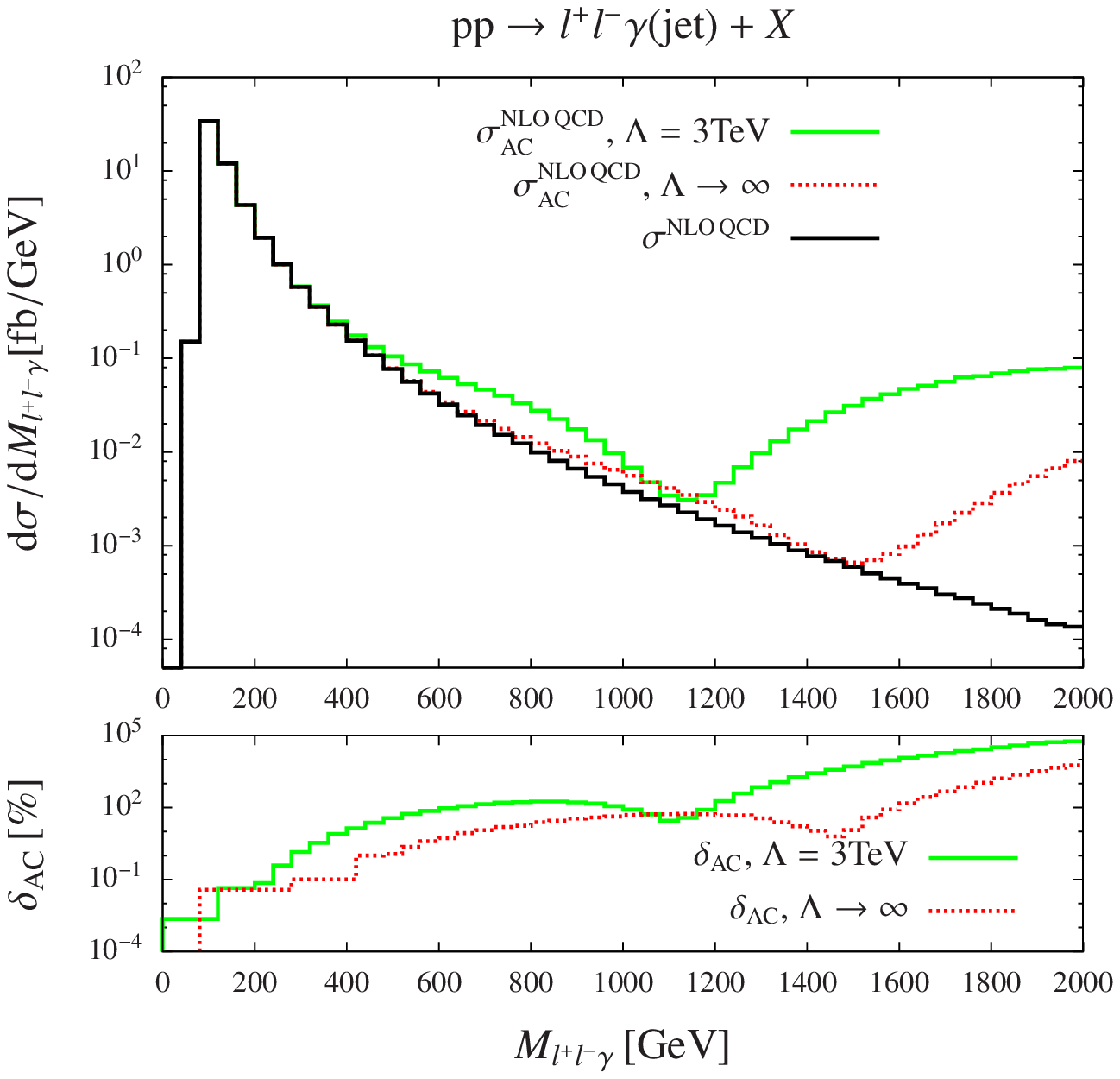}
 }
\caption{\label{fi:minv_zll_ac} Absolute and relative contributions of
  aTGCs to the invariant-mass distribution of the charged leptons
  (left) and to the invariant three-body mass of the charged leptons
  and the hardest photon (right).}
\end{figure}

Next we analyse the invariant mass of the charged leptons and the invariant three-body 
mass of the charged leptons and the photon shown in \reffi{fi:minv_zll_ac}. 
Starting with the $M_{\Plp\Plm}$ distribution we see that the aTGCs
only have a significant impact on the invariant-mass distribution
around the $\PZ$~pole. At higher invariant masses up to several
$100\GeV$ the aTGCs have almost no effect. This can be explained
exactly in the same way as in the case of $\PW+\gamma$ production,
where amongst others we analysed the impact of aTGCs on the
transverse-mass distribution of the charged lepton and the neutrino in
\citere{Denner:2014bna}. At large invariant mass $M_{\Plp\Plm}$ the
intermediate bosons $\PZ$ and $V$ coupled to the anomalous
$V\gamma\PZ$ ($V=\gamma,\PZ$) vertex are far off shell. This fact
allows us to explain the small effect of aTGCs at large invariant
masses, which are typically driven by disturbing the unitarity
cancellations of the SM amplitude. In case of resonant $\PZ$~bosons
these cancellations occur for longitudinally polarized $\PZ$~bosons
with momentum $q^\mu$ and virtuality $q^2\sim\MZ^2$, where the
effective $\PZ$~polarization vector behaves like
$\varepsilon_{\mathrm{L}}^\mu\sim q^\mu/\sqrt{q^2}\sim q^\mu/\MZ$.
For large invariant masses, the $\PZ$~virtuality is large,
$q^2\gg\MZ^2$, so that $\varepsilon_{\mathrm{L}}^\mu\sim
q^\mu/\sqrt{q^2}$ is suppressed, and no large cancellations are
necessary within the amplitude to avoid unitarity violations in the
SM. The suppression in the polarization
$\varepsilon^{\mu}_{\mathrm{L}}$ explains why there is no effect of
the aTGCs visible in the high-mass tail of the invariant-mass
distribution in contrast to other scale-dependent distributions.  The
impact of aTGCs near the Z~pole is much higher for the case without
form factor in comparison to the setup with $\Lambda=3\TeV$, although
the AC values for $\Lambda\to\infty$ are much smaller.  The arguments
given above for the transverse-momentum distributions, which lead to
the expectation that the AC effects for $\Lambda\to\infty$ should be
smaller, do not apply here, because the aTGC effects without form
factor are dominated by extremely large scattering energies even for
$M_{l^+l^-}\sim\MZ$. Here, it should be kept in mind that the limits
\refeq{eq:AC} were obtained for LHC energies of $7/8\TeV$, but our
results are for an energy of $14\TeV$.

In \reffi{fi:minv_zll_ac} (right) we observe a large impact of the
aTGCs on the distribution in the invariant three-body mass. The
relative corrections from the aTGCs obtained with and without a form
factor grow to $10^{3}$ and $10^{2}$ at $2\TeV$, respectively. With
the same arguments as before we can now explain why the effect of the
aTGCs is so large here. A high invariant three-body mass can occur
while the outgoing $\PZ$~boson is on shell if the outgoing photon
carries away a sufficiently large amount of the energy brought into
the $V\gamma\PZ$ vertex ($V=\gamma,\PZ$) by the incoming boson.
Therefore the longitudinal polarization vector of the outgoing
$\PZ$~boson, which is effectively produced by the leptonic decay
current, is not suppressed leading to a large contribution of the
aTGCs at high invariant three-body masses.  In view of the hierarchy
of the impact of aTGCs in our two setups, the arguments given for the
transverse-momentum spectra again apply, i.e.\ the case without form
factor shows a smaller AC impact up to moderate scales, because the
corresponding set of AC values is stronger constrained by data in this
range.

\subsubsection{$\Pp\Pp\to \bar{\nu} \nu \gamma + X$}

\begin{figure}     
 \centerline{
 \hspace{-3em}
         \includegraphics[scale=0.6]{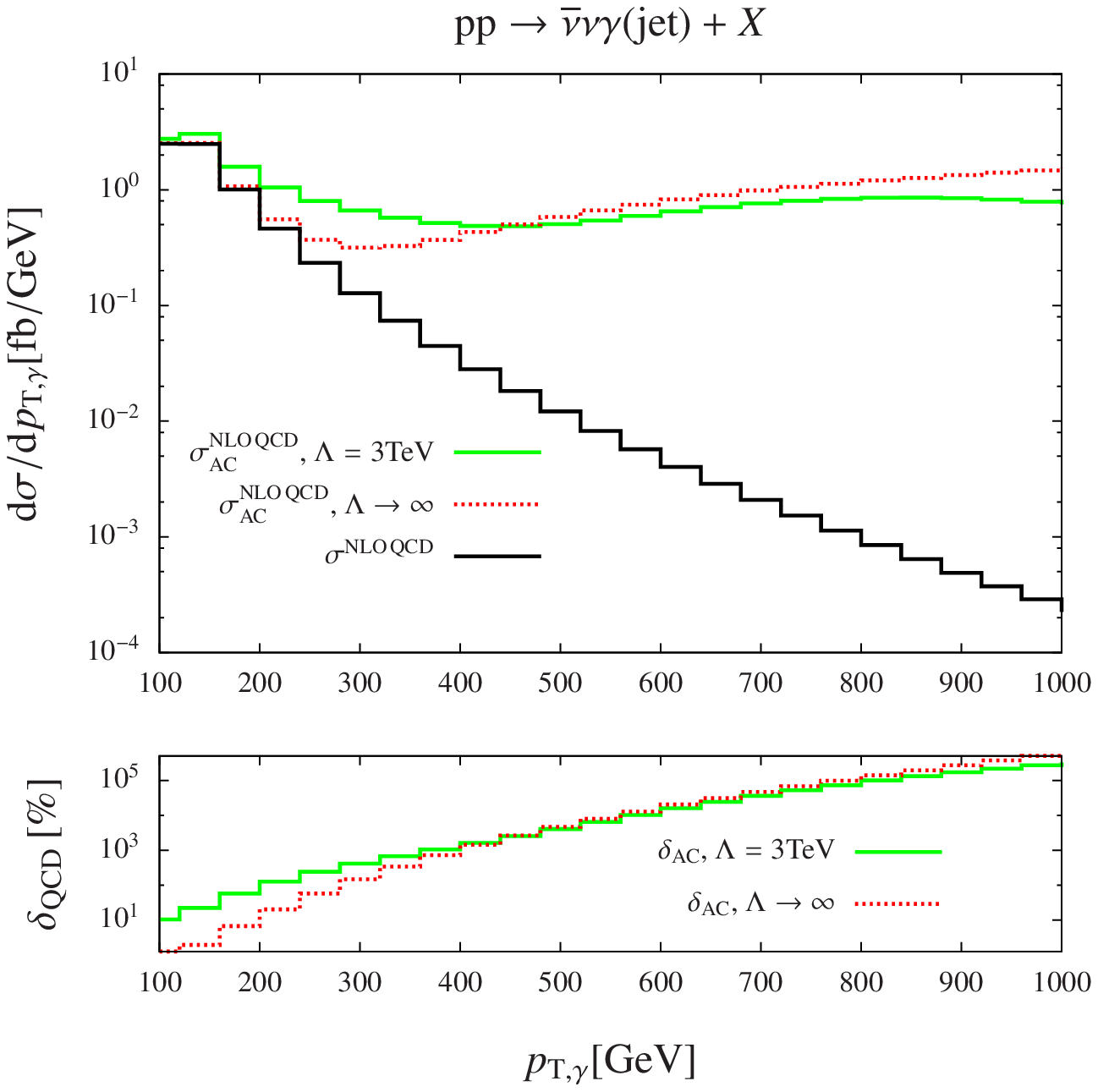}
 \hspace{-3em}
         \includegraphics[scale=0.6]{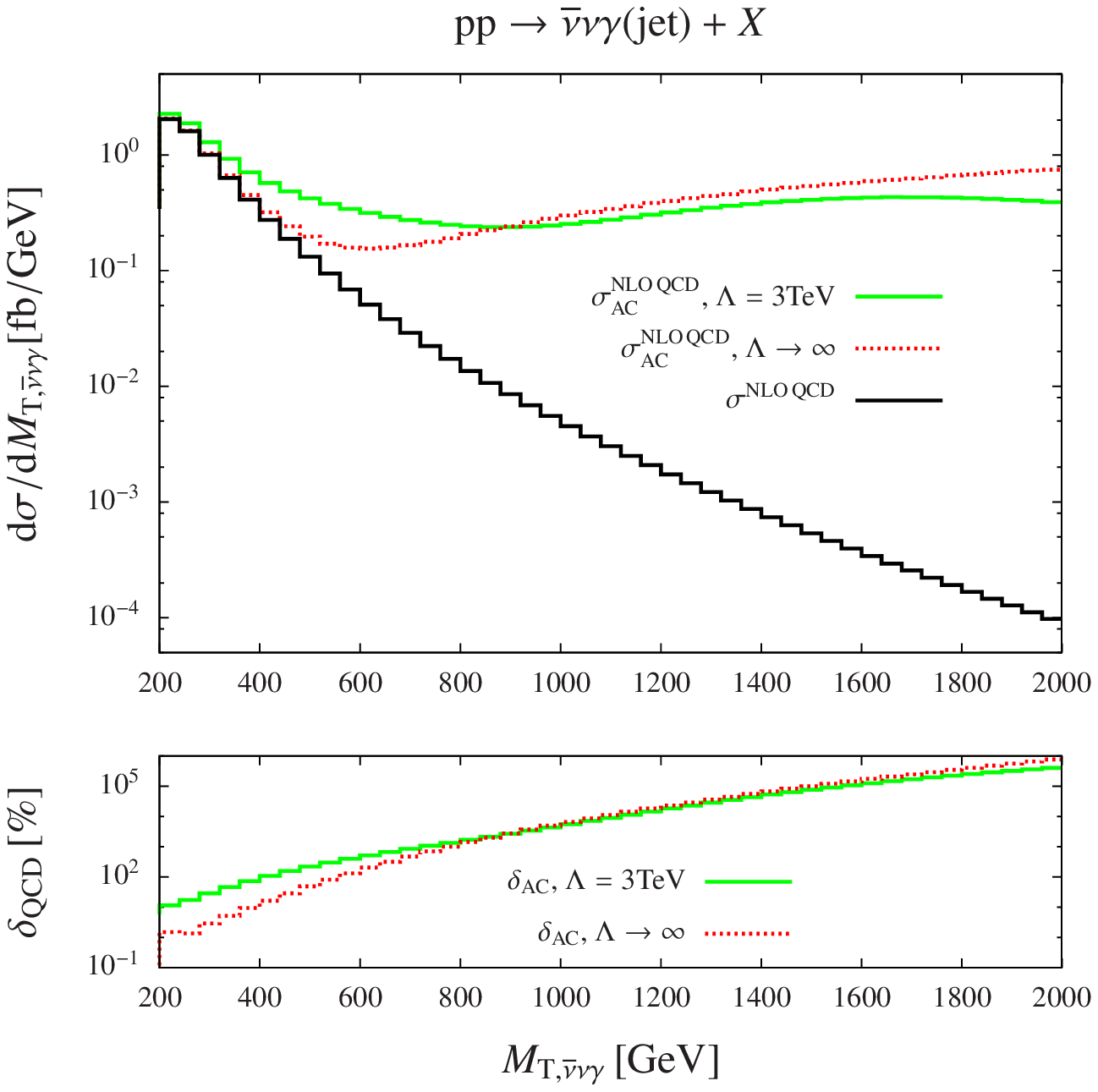}
 }
\caption{\label{fi:pt_mt_znn_ac} Absolute and relative contributions
  of aTGCs to the transverse-momentum distribution of the photon
  (left) and to the transverse three-body mass of the neutrinos and
  the hardest photon (right).}
\end{figure}

Turning 
to the invisible decay channel of the $\PZ$ boson we show the transverse-momentum 
distribution of the photon and the transverse three-body mass of the neutrinos and the
photon in \reffi{fi:pt_mt_znn_ac}. 
The $p_{\mr{T}}$ distribution of the photon 
receives the same corrections from aTGCs as the $p_{\mr{T}}$ distribution of the photon
in case of the visible decay channel of the $\PZ$~boson. This is due to the fact that the
cross-section contributions by
aTGCs do not depend on the FS particles. 
In case of the {\it transverse} three-body mass distribution the relative contribution of the aTGCs
increases much faster than in the $M_{\Plp\Plm\gamma}$ distribution, 
which is again due
to the influence of events where the three-body invariant mass is much higher than
the transverse three-body mass.
Since aTGC contributions grow with higher invariant masses, 
the relative contribution of the aTGCs increases faster in case of the transverse three-body mass.  

\section{Conclusions}
\label{se:concl}

Analyzing $\PZ + \gamma$ final states at hadron colliders offers
several directions to probe the SM and to look for traces of new
physics.  Final states with charged leptons, $l^+l^-\gamma$, are ideal
to look for non-standard effects in $\PZ\PZ\gamma$ and
$\PZ\gamma\gamma$ couplings, which do not exist in the SM as
elementary interactions.  On the other hand, final states with
invisibly decaying Z~bosons and a hard photon, known as mono-photon
production, are prominent signatures of many exotic new-physics
models.  Both types of reactions require improved theoretical
predictions for experimental analyses at run~2 of the LHC.

In this paper we have improved the knowledge of $\PZ + \gamma$
production on the side of electroweak higher-order corrections for
both process types.  Specifically, we have calculated the full
next-to-leading-order electroweak corrections to the processes
$\Pp\Pp\to l^+l^-/\bar\nu\nu+\gamma+X$, taking into account all
off-shell effects of the Z~boson using the complex-mass scheme and
including all partonic channels ($\Pq\gamma$ and $\gamma\gamma$) with
initial-state photons.  In order to discuss the phenomenological
separation of $\PZ+\gamma$ or $\PZ+\mathrm{jet}$ production, we have
recalculated the NLO QCD corrections.  The actual distinction between
hard photons and hard jets in their overlap region is
performed in two alternative ways by employing a quark-to-photon
fragmentation function or Frixione's cone isolation.

Reflecting the known general feature of EW corrections in the TeV
range, we find those corrections of the size of several $10\%$ in
distributions, while their impact on integrated cross sections remains
at the level of some percent.  The impact of photon-induced channels
is moderate or small throughout, reaching some percent in extreme
regions of distributions. 

\change{We estimate the theoretical uncertainties from missing 
  higher-order electroweak corrections to be of the order of $0.5\%$ for
  integrated cross sections and $1\%$ for differential distributions.
  For distributions, where the contributions from photon-induced
  channels exceed one percent these contributions should be viewed as
  an additional theoretical uncertainty. Moreover, if electroweak
  corrections surpass $10\%$ their square should be considered as a
  measure for the missing electroweak corrections beyond next-to-leading 
  order.}

On top of our complete NLO EW+QCD predictions in the SM, we have
included non-standard effects in $\PZ\PZ\gamma$ and $\PZ\gamma\gamma$
couplings at the NLO QCD level in the usual approach of anomalous
couplings, on which previous Tevatron and LHC analyses were based.  In
view of future global analyses of non-standard couplings in the
effective field theory approach with dimension-six operators, also the
$\PZ + \gamma$ analyses should be performed in this framework.  On the
theoretical side this task is straightforward at the NLO QCD level, but
delicate if EW corrections should be combined with non-standard
operators beyond a mere addition. These issue is left to future work.

Within the SM, our calculations constitute an important part of
state-of-the-art predictions for $\PZ+\gamma$ production.
To this end, our results should be combined with the recently
published next-to-next-to-leading-order QCD predictions,
e.g.\ upon including differential reweighting factors for the
EW corrections on top of the absolute QCD predictions.
This combination
should provide the necessary precision in predictions required for the
coming data analysis at the LHC at its design energy and luminosity.

\subsection*{Acknowledgements}

The work of S.D.\ and M.H.\ is supported by the German Research Foundation (DFG) via
grant DI 784/2-1 and the Research Training Group
GRK 1102 ``Physics at Hadron Colliders''.
The work of A.D.\ and C.P.\ is supported by the Research Training Group
GRK 1147 ``Theoretical Astrophysics and Particle Physics''.

\bibliography{Bibliography}

\end{document}